\documentclass[graybox, secnum]{svmult}

\usepackage[square,numbers]{natbib}
\usepackage{mathptmx}       
\usepackage{helvet}         
\usepackage{courier}        
\usepackage{type1cm}        
\usepackage{amssymb} 

\usepackage{makeidx}         
\usepackage{graphicx}        
\usepackage{multicol}        
\usepackage[bottom]{footmisc}
\usepackage{hyperref}        
\usepackage{soul}            
\hypersetup{colorlinks=true,urlcolor=blue}
\makeindex             
\newcommand{\apj}{Astrophys J }
\newcommand{\apjl}{Astrophys J Lett }
\newcommand{\aj}{Astronom J }
\newcommand{\apjs}{Astrophys J Suppl }
\newcommand{\mnras}{MNRAS }
\newcommand{\aap}{Astron Astrophys }

\newcommand{\aapr}{Astron Astrophys Rev }
\newcommand{\nat}{Nature }

\newcommand{\pasp}{Publications Astron Soc Pacific }
\newcommand{\araa}{Annu Rev Astron Astr }
\newcommand{\na}{New Astr }
\newcommand{\alphaox}{\alpha_{\rm ox}}
\newcommand{\lx}{\rm L_{2 \;{\rm keV}}}
\newcommand{\angstrom}{\mbox{\normalfont\AA}}

\begin{document}
\title*{Active Galactic Nuclei and their demography through cosmic time}
\author{Stefano Bianchi\thanks{corresponding author} and Vincenzo Mainieri\thanks{corresponding author} and Paolo Padovani\thanks{corresponding author} }
\institute{Stefano Bianchi \at  Dipartimento di Matematica e Fisica, Universit\`a degli Studi Roma Tre, via della Vasca Navale 84, I-00146 Roma, Italy, \email{bianchi@fis.uniroma3.it}
\and Vincenzo Mainieri \at European Southern Observatory, Karl-Schwarzschild-Str. 
2, D-85748 Garching bei M\"unchen, Germany, \email{vmainier@eso.org}
\and Paolo Padovani \at European Southern Observatory, Karl-Schwarzschild-Str. 
2, D-85748 Garching bei M\"unchen, Germany, \email{ppadovan@eso.org}}
%
%
\maketitle
\abstract{Active Galactic Nuclei (AGN) are highly energetic astrophysical sources 
powered by accretion onto supermassive black holes in galaxies, which present unique 
observational signatures covering the full electromagnetic spectrum (and more) over 
about twenty orders of magnitude in frequency. We first review the main AGN properties 
and diversities and show that they can be explained by a small number of parameters. 
We then discuss the so-called Unification Models for non-jetted AGN, according to which 
these sources are believed to have the same nuclear engine and circumnuclear matter, with 
the same geometry for the obscuring structure. This simplified scenario, however, cannot 
explain all the observed complexities, such as the presence of multiple absorbers on 
different physical scales, including recent X-ray observations of circumnuclear matter.
Finally, we touch upon AGN evolution in the X-ray and $\gamma$-ray bands.}


\section{Active Galactic Nuclei as multi-wavelength and multi-messenger emitters}\label{sec:AGN_emitters}

The discovery of quasars \citep{Schmidt_1963} revolutionized astronomy, 
opening up at the same time a whole new field, that of 
AGN \citep[e.g.][for historical 
details]{Donofrio_2012,Kellermann_2015}. 
As implicit in their name, AGN outshine the nuclei of ``normal'' galaxies. 
Their emission is unrelated to the nuclear fusion powering stars and is 
now instead universally accepted to be connected to an actively accreting 
central supermassive ($\gtrsim 10^6~M_{\odot}$) black hole (SMBH). 
This leads to a host of interesting properties, which include: 
(1) very high powers (up to a bolometric luminosity $L_{\rm bol} \approx 10^{48}$ erg s$^{-1}$), which make AGN the most powerful, non-explosive sources in the Universe,  visible up to very high redshifts \citep[currently $z_{\rm max}=7.642$:][]{Wang_2021}; 
(2) relatively small (milliparsec-size) emitting regions 
as inferred from their variability  \citep[e.g.][]{Ulrich_1997}, 
which imply very high energy densities; (3) very strong evolution of 
their luminosity functions \citep[LFs; e.g.][]{Merloni_2013,Brandt_2015}, which give the
number of sources per luminosity bin and per unit volume (see Sect. \ref{sec:dem}); 
(4) detectable emission covering the whole electromagnetic spectrum 
\citep{Padovani_2017b}. 

\begin{figure}
\centering
\hspace{-1.18truecm}
\includegraphics[width=12.0cm]{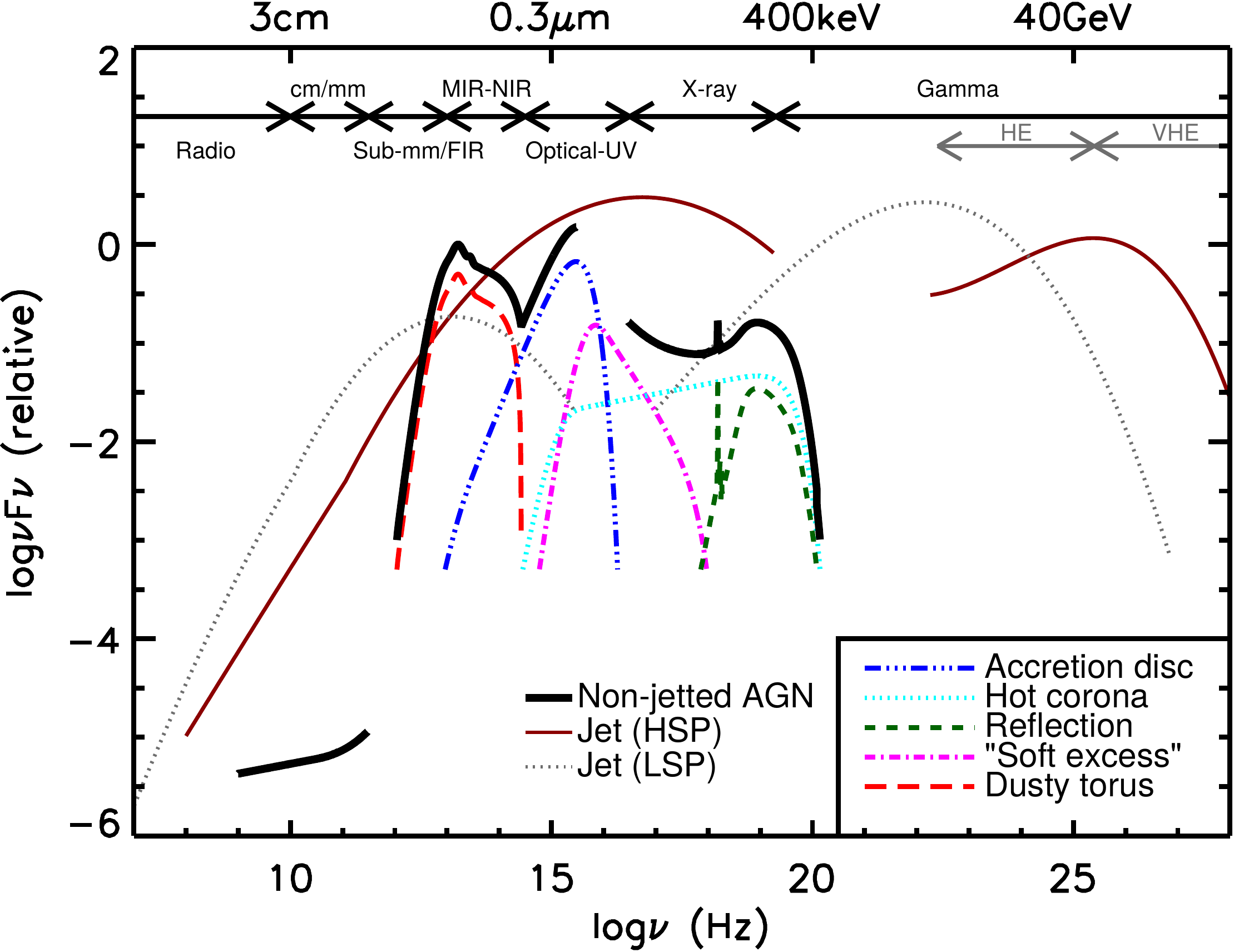}
\caption{A schematic representation of the spectral energy distribution 
(SED) of radiatively efficient AGN, loosely based on the observed SEDs of 
non-jetted quasars 
\citep[e.g.][]{Elvis_1994,Richards_2006}. The black solid curve represents 
the total emission and the various coloured curves (shifted down for clarity) 
represent the individual components. The jet 
SED is also shown for a high synchrotron peaked blazar (HBL; based on the SED of 
Mrk 421) and a low synchrotron peaked blazar (LBL; based on the SED of 3C 454.3). 
The various AGN components and classes are described in details later in the text. 
Figure reproduced from \cite{Padovani_2017b}, Fig. 1, with permission. 
Image credit: C. M. Harrison.}
\label{fig:SED}       
\end{figure}

\begin{figure}
\centering
\hspace{-1.18truecm}
\includegraphics[width=12.0cm]{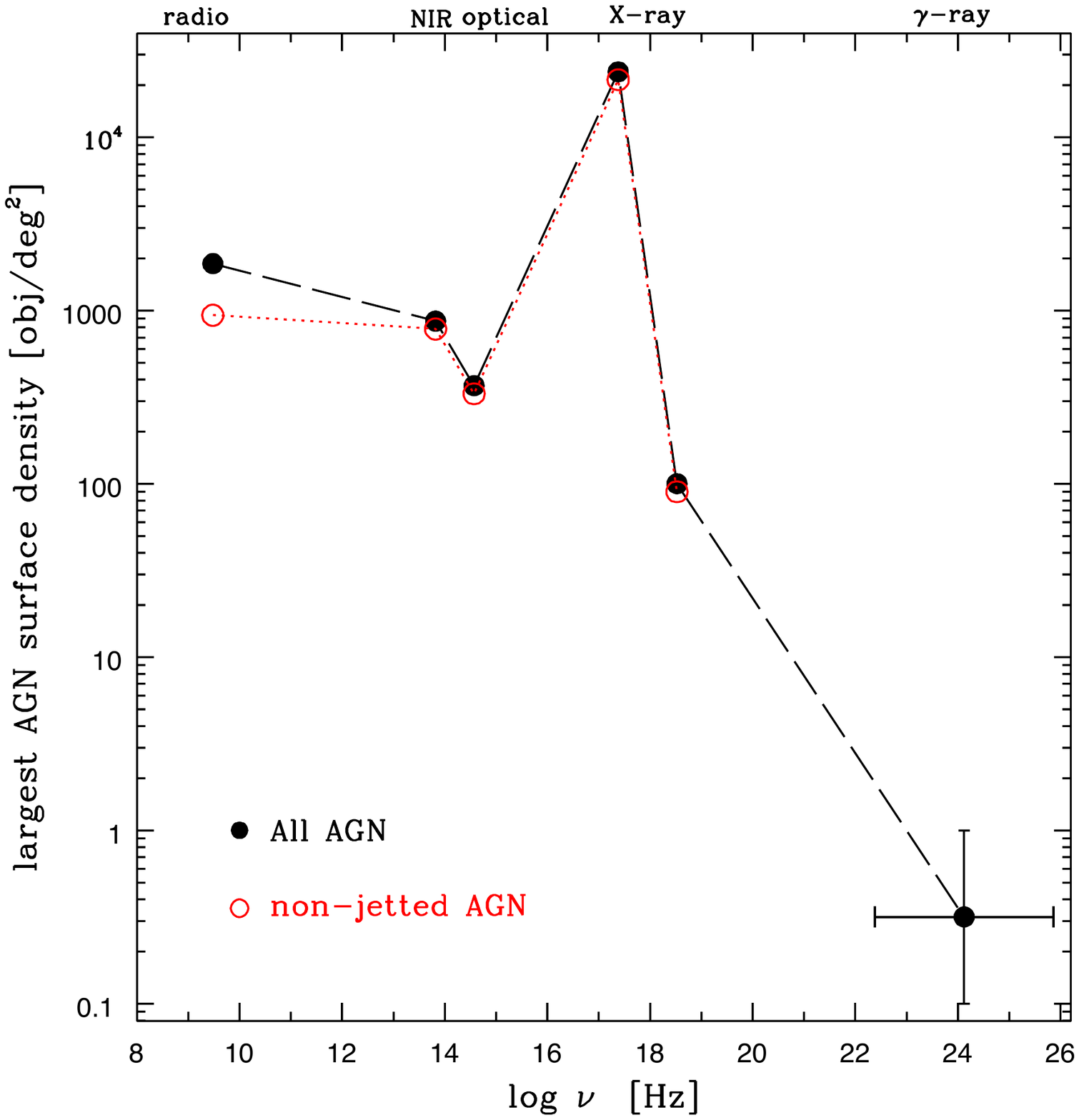}
\vspace{-3.5truecm}
\caption{The largest observed AGN surface density by surveys over the
  whole electromagnetic spectrum. Black filled points refer to all AGN,
  open red points are for non-jetted AGN. The latter are actually directly measured
  only in the radio band, while in the NIR to X-ray bands they have been
  derived by simply subtracting $10\%$ from the total values. Non-jetted AGN
  have not been detected in the $\gamma$-ray band. Figure updated from \cite{Padovani_2016}, Fig. 11,
  and \cite{Padovani_2017b}, Fig. 21, where one can find the references to the relevant samples 
  and facilities, to include new radio \citep[][and V. Smol{\v c}i{\'c}, private 
  communication]{Delvecchio_2017} and $\gamma$-ray results \citep{Marcotulli_2020}. The $\gamma$-ray 
  point is the mean of the two values
  shown in Fig. 6, bottom, of the latter reference, which refer to photon and energy fluxes.}
\label{fig:densities}       
\end{figure}

The last property means that AGN have been, and are being, discovered 
in {\it all} spectral bands, by employing a variety of methods and 
selection techniques. The crucial point is that {\it different wavelengths 
provide different windows} on AGN physics. Namely, as discussed extensively 
in \cite{Padovani_2017b}, the infrared (IR) band is 
mostly sensitive to obscuring material and dust, the optical/ultraviolet (UV) 
band is related to emission from the accretion disk (the so-called ``big blue bump''), 
while the X-ray band 
traces Comptonized emission from a hot corona. $\gamma$-ray and (high flux density) 
radio samples, instead, preferentially select AGN emitting strong 
non-thermal radiation coming from relativistic jets, i.e., streams of plasma
with speeds approaching the speed of light, and associated lobes (see Fig. 
\ref{fig:SED}). 
As a result, the observed surface densities of AGN (i.e., the number of sources 
per square degree) also vary strongly across the 
electromagnetic spectrum due to a complicated mix of technological limitations, 
selection effects, and physical processes, as shown by Fig. 
\ref{fig:densities}.  

The broad-band AGN spectrum shown in Fig. \ref{fig:SED} shows that there are two regions which are dominated by the accretion power: the X-ray and optical-UV. It is not surprising therefore that, since the earliest studies on AGN, there has been particular attention to the luminosity ratio of these two bands \citep[e.g.][]{tananbaum79} parameterized by the so-called optical to X-ray spectral index ($\alpha_{\rm ox}$) defined as:
\begin{equation}
\alphaox=-\frac{Log[\lx/L_{2500 \angstrom}]}{Log[\nu_{\rm 2 keV}/\nu_{\rm 2500 \angstrom}]}=-\frac{Log[\lx/L_{2500 \angstrom}]}{2.605}
\end{equation}

The study of the luminosity dependence of $\alpha_{\rm ox}$ has been used to understand the relation between the disk emission and the coronal emission, respectively traced in the UV and X-ray bands. We refer the reader to Chap. 2 for a detailed discussion on the use of $\alpha_{\rm ox}$ to trace UV and X-ray emission and to Chap. 4 for cosmological quantities.

\subsection{The AGN ``zoo''}\label{sec:AGN_zoo}

\setcounter{footnote}{0}

The way AGN have been selected has led, almost unavoidably, to an explosion
of AGN classes and sub-classes (see Tab. 1 of \cite{Padovani_2017b}), 
which outsiders to the field, but sometimes insiders as well, find confusing, 
to say the least (ask physicists!). Luckily, reality is much simpler,
with most, if not all, of these apparently different classes being due to 
changes in a small number of parameters, namely: orientation (of the obscuring material
and of the jet) 
\citep[e.g.][]{Antonucci1993,Urry_1995,Netzer_2015}, radiative efficiency (see further on), 
related to the accretion rate and therefore the Eddington ratio ($L/L_{\rm Edd}$), 
\citep[e.g.][]{Heckman_2014}, the presence (or absence) of strong 
relativistic jets \citep[e.g.][]{Padovani_2016,Padovani_2017a,Radcliffe_2021}, 
and possibly the host galaxy and the environment. $L/L_{\rm Edd}$ 
is the ratio between the observed luminosity and the Eddington luminosity, 
$L_{\rm Edd} = 1.3 \times
  10^{46}~(M/10^8 \rm M_{\odot})$ erg s$^{-1}$, where $\rm M_{\odot}$ is one solar
  mass, which is the maximum isotropic luminosity a body can achieve when there is
  balance between radiation pressure (on the electrons) and gravitational
  force (on the protons).

\begin{figure}
\centering
\includegraphics[width=12.0cm]{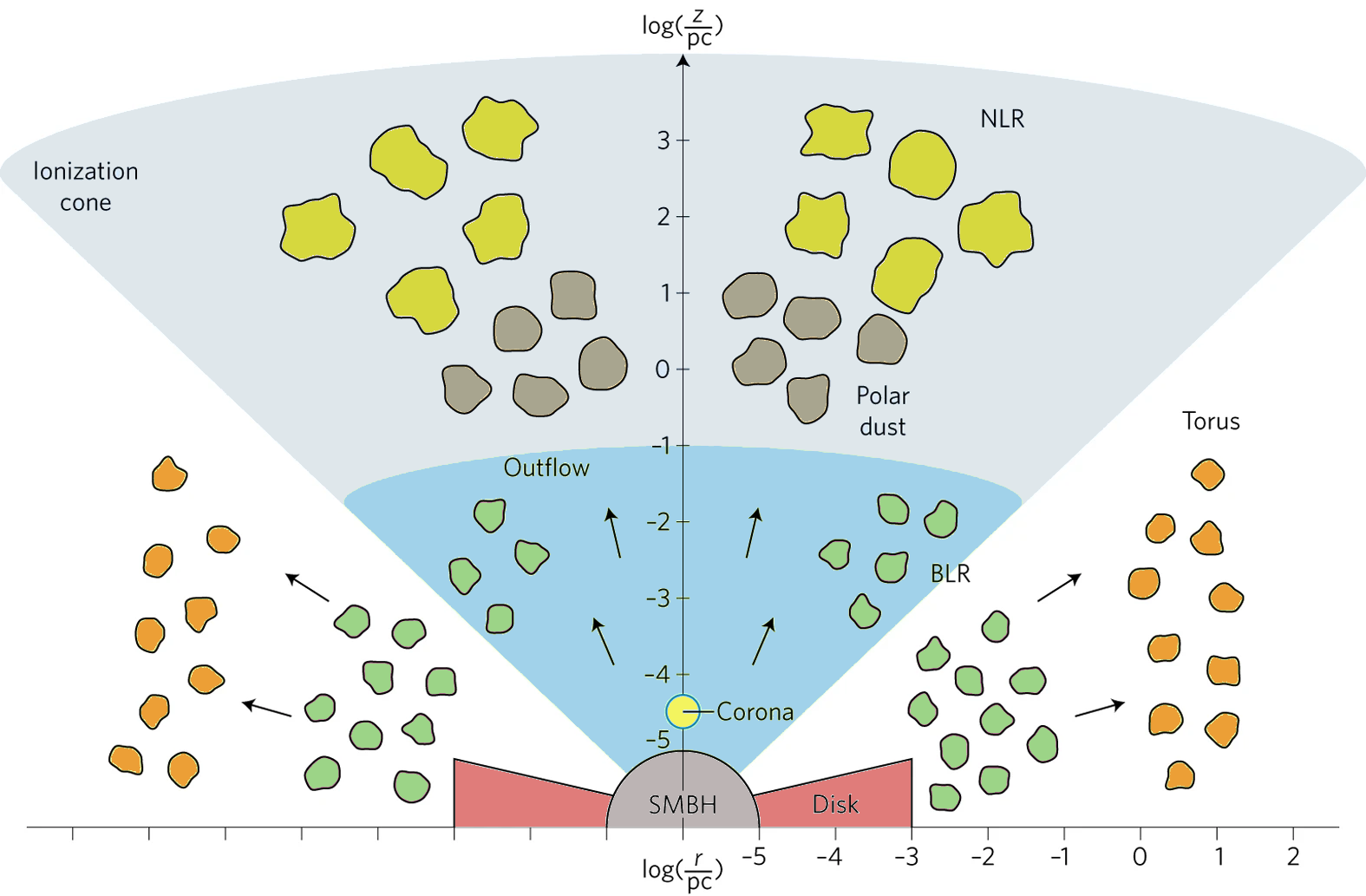}
\caption{A schematic view of the inner parts of a non-jetted AGN indicating, from the centre to host-galaxy scales, the SMBH, the accretion disk and the corona, the BLR, the torus, and the NLR. Different colours indicate different compositions or densities. Figure adapted from \citet{RamosAlmeida2017}, courtesy of C. Ramos Almeida and C. Ricci. }
\label{fig:sketch}       
\end{figure}

Fig. \ref{fig:sketch} provides a schematic view of the inner parts of a non-jetted AGN and highlights, from the centre to host-galaxy scales, the SMBH, the accretion disk and the corona, the broad line region (BLR), the torus, and the narrow line region (NLR). X-rays are produced via Comptonization of the accretion disk's photons by a corona of hot electrons. The resulting power law spectrum interacts with the surrounding matter, being absorbed (for example by the torus) or `reprocessed' (by the accretion disk itself, the torus, the NLR), giving rise to further spectral components, in particular a Compton reflection peaking at 20-30 keV, and fluorescent emission lines, the most prominent being the K$\alpha$ line from neutral iron at 6.4 keV. 

Turning to the various classes,
radio quasars are radiatively efficient (see below), jetted AGN, seen face-on. 
The fact that they display strong and Doppler broadened lines in their 
spectra (with full-width half maximum [FWHM] $> 1000$ km/s), in fact, 
implies that we are looking inside the torus, i.e. 
dust in a flattened configuration roughly perpendicular to the jet. 
Their edge-on versions are high-excitation (i.e. radiatively efficient) radio galaxies, 
most of which are of the
Fanaroff-Riley (FR)\footnote{\cite {Fanaroff_1974} divided radio
galaxies into type I and II by using the ratio of the separation of the 
highest surface brightness regions on opposite sides of the central 
galaxy and the extent of the source measured from the lowest 
surface brightness contour, with FR~Is and FR~IIs having a ratio 
$<0.5$ and $>0.5$, respectively.} II type. In this case 
the central nucleus and surrounding material (including the broad line 
emitting clouds) are obscured by the dust and only narrow lines, emitted
by more distant clouds, are visible in their optical
spectra (Fig. \ref{fig:sketch}). Seyfert 1s\footnote{It is now generally agreed upon that the
  distinction at $M_{\rm B} \sim -23$ used in the past to separate quasars
  and Seyferts, or stellar and non-stellar (i.e., extended) sources, is not
  a physical one, and that Seyfert 1's are lower
  luminosity versions of quasars. Still, the absolute magnitude cut might
  be useful to roughly separate AGN fainter and brighter
  than the brightest galaxies \citep[e.g.,][]{Condon_2013}.} and quasars are similarly efficient and oriented but are 
  non-jetted AGN, whereby Seyfert 2's and
  type 2 quasars are their edge-on relatives. Turning to radiatively inefficient
  AGN, BL Lacertae objects (BL Lacs) represent the jetted, face-on kind, while the edge-on version include 
  low-excitation radio galaxies, most of which are 
  FR Is. Finally, non-jetted, inefficient AGN are collectively
  lumped under the Low-ionization Nuclear Emission-line Region (LINER) class\footnote{Note however
  that not all LINERs appear to be ionized by AGN: \cite[e.g.][]{Sarzi2010}.}. Non-jetted AGN make up the majority ($> 90\%$) of all AGN. 
  
  Radiatively inefficient accretion is related to low $L/L_{\rm Edd}$ ($\lesssim 0.01$), and therefore typically also low luminosities, while 
  radiatively efficient sources generally have $L/L_{\rm Edd} \gtrsim 0.01$ \citep[e.g.][]{Heckman_2014}. 
  The difference in $L/L_{\rm Edd}$ is generally associated with 
  a switch between a standard accretion, i.e. radiatively efficient, geometrically thin (but optically thick) disk 
  accretion flow and a radiatively inefficient, geometrically thick (but optically thin) disk accretion flow 
  \citep[e.g.][and references therein; see also Sect. \ref{sec:subl_r}]{Feng_2014,Padovani_2017b}. We note that the SED shown 
  in Fig. \ref{fig:SED} is representative of radiatively efficient AGN. The SEDs of 
  low-luminosity, radiatively inefficient AGN are somewhat different. Namely, when normalised
  at the same X-ray power, they show a somewhat higher radio power (by about one order
  of magnitude) and a much weaker big blue bump in the UV \citep[e.g.][]{Younes_2012}. 
  
\begin{figure}
\centering
\hspace{-1.18truecm}
\includegraphics[width=12.0cm]{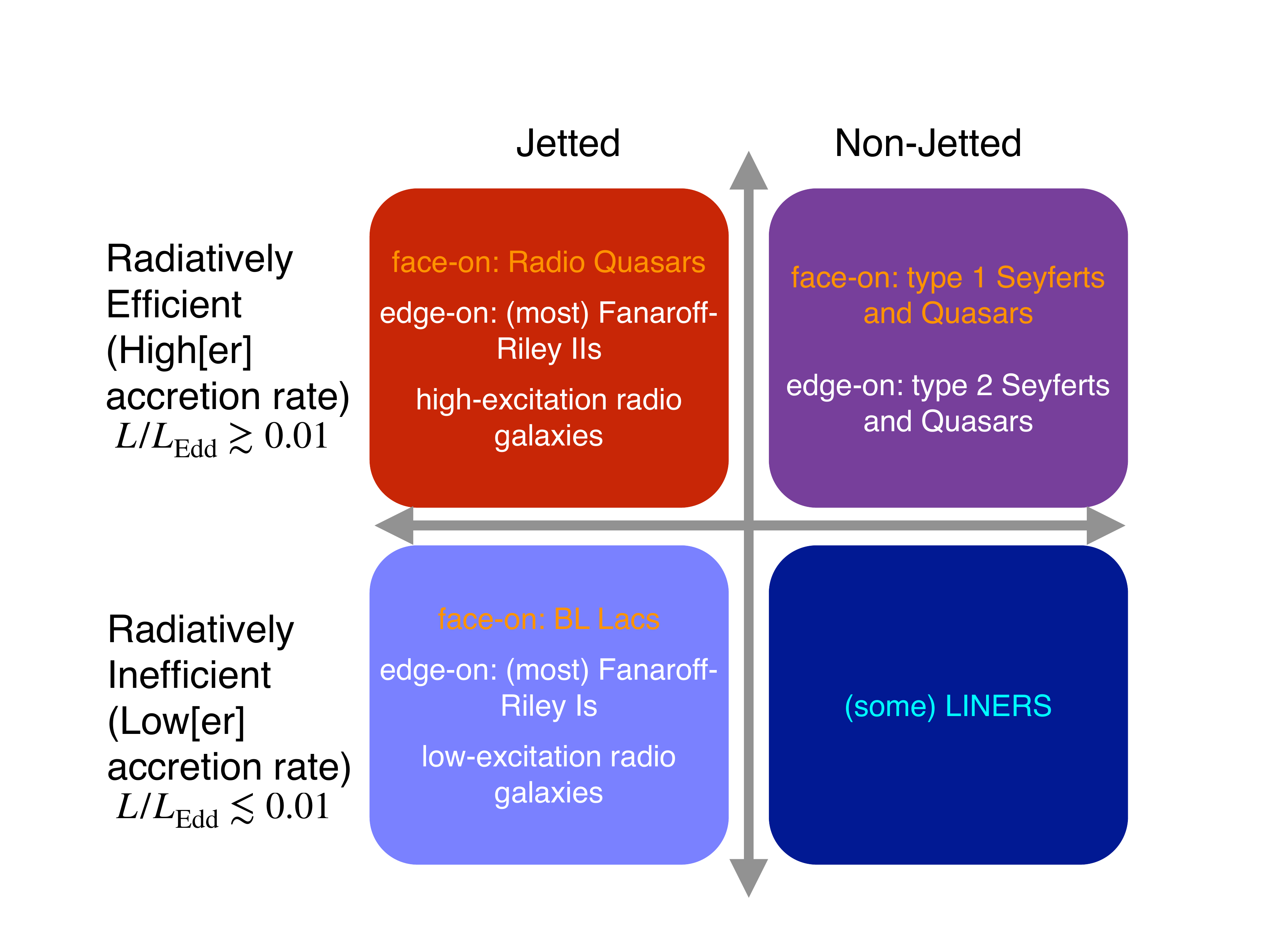}
\caption{A ``minimalistic'' approach to AGN classification using three parameters: 
radiative efficiency, related to $L/L_{\rm Edd}$, relativistic jet presence, 
and orientation, showing the classes associated with each broad range in 
this parameter space. Figure adapted from \cite{Padovani_2017b}, Fig. 22. 
Originally adapted by R. Hickox from schematic by P. Hopkins.}
\label{fig:scheme}       
\end{figure}

  Blazars are simply jetted AGN oriented 
  at very small angles ($\lesssim 15 - 20^{\circ}$) with respect to the line of sight and  include two main classes, of two 
  opposite accretion types: BL Lacs (radiatively inefficient) and flat-spectrum radio quasars (FSRQs: radiatively 
  efficient). As a result, the former have often totally featureless optical spectra, sometimes exhibiting weak 
  absorption and emission lines, while the latter display broad, quasar-like 
  strong emission lines \citep{Urry_1995,Padovani_2017b}.  The SED of blazars is
  characterised by a typical ``double hump'' shape (in $\nu - \nu f(\nu)$ space: see Fig. \ref{fig:SED}).
  The low-energy component, peaking between the IR and the X-ray band, is generally attributed to synchrotron 
  radiation produced by relativistic electrons in a magnetic field. Objects where this component peaks at low
  energies ($\nu_{\rm synch~peak} <10^{14}$~Hz [$<$ 0.41 eV]) are called LBLs/LSPs, while objects peaking
  at high energies ($\nu_{\rm synch~peak} > 10^{15}$~Hz [$>$ 4.1 eV] are named HBLs/HSPs
  \citep{Padovani_1995,Abdo_2010}. Intermediate sources are the IBLs/ISPs. The nature of the second component, 
  that extends well into the $\gamma$-ray band, is still debated (Sect. \ref{sec:AGN_HE}).

Fig. \ref{fig:scheme} shows 
that a ``minimalistic'' approach to AGN classification using just 
the three parameters discussed above can explain the most important AGN classes. 

\subsection{AGN as high-energy and multi-messenger sources}\label{sec:AGN_HE}

The feature which is most relevant to this handbook is the fact that AGN 
are very powerful high-energy, i.e. $E \gtrsim 1$ keV, emitters. The X-ray 
and $\gamma$-ray bands, however, sample somewhat different AGN populations.
X-ray emission appears to be a universal AGN feature, which implies that
both jetted and non-jetted AGN are strong\footnote{The oft-quoted value of 
$L_{\rm 2 - 10~keV} \gtrsim 10^{42}$ erg s$^{-1}$ is used to separate AGN from
star-forming galaxies. However, as pointed out by \cite{Padovani_2016}, 
plenty of AGN, especially, but not only, of the radiatively inefficient 
jetted type, have $L_{\rm x}$ below this value.} X-ray emitters. This,
together with the sensitivity of X-ray observatories, explains why 
the observed AGN surface density is highest in the X-ray band (Fig. \ref{fig:densities}).   
However, {\it only} jetted AGN, almost all of the blazar type, 
manage to emit in the $\gamma$-ray ($E \gtrsim 1$ 
MeV) regime, as non-jetted AGN are not detected by {\it Fermi} \citep{4FGL}\footnote{
To be more precise, a handful of Seyferts, including NGC 1068 and NGC 4945,
are listed in the {\it Fermi-}4FGL-DR3 catalogue \citep{4FGL-DR3} but their $\gamma$-ray 
emission is thought to be related to starburst emission in their host galaxy and not to their
SMBH \citep{Ajello_2020}.}, the $\gamma$-ray space mission surveying the high-energy sky in the 
$\sim$ 50 MeV -- 1 TeV band\footnote{\url{https://fermi.gsfc.nasa.gov/}}. In fact, blazars make up $\sim 60\%$ of the {\it Fermi}  sky and the vast majority of extragalactic $\gamma$-ray sources 
\citep{4FGL-DR3}, while 92\% of 
extragalactic objects above 1 TeV\footnote{\url{http://tevcat.uchicago.edu}} are also blazars, studied 
mostly with Imaging Atmospheric Cherenkov Telescopes (IACTs) such as MAGIC\footnote{\url{https://magic.mpp.mpg.de/}}, H.E.S.S.\footnote{\url{https://www.mpi-hd.mpg.de/hfm/HESS/}}, and 
VERITAS\footnote{\url{https://veritas.sao.arizona.edu/}}. IACTs, unlike {\it Fermi}, have a relatively small field of view and observe from 
the ground the particle showers produced by the impact of very high-energy ($\sim$ 50 GeV 
-- 10 TeV) $\gamma$-ray photons on the top layers of the atmosphere, either through the 
Cherenkov light they generate, or via the direct detection of the charged particles in the 
shower.

AGN outflows, i.e. large-scale winds of matter driven by the central SMBH, were also predicted to be (faint) $\gamma$-ray sources,
as their semi-relativistic speeds (up to $\sim$ 50,000 km s$^{-1}$) can drive a shock that 
accelerates and sweeps up matter \citep[e.g.][]{King_2015} (see also Sect. \ref{sec:subl_r}). The protons accelerated by these shocks 
can then generate low-level $\gamma$-ray emission via collisions with protons in the interstellar 
medium \citep[e.g.][]{Lamastra_2017,Liu_2018}. This has been detected very recently in AGN 
with ultra-fast outflows (UFOs) through a stacking analysis by {\it Fermi} \cite{Fermi_2021}.

Blazars have also recently entered the multi-messenger arena. A high-energy neutrino event 
($E \sim 290$ TeV) observed by IceCube\footnote{\url{http://icecube.wisc.edu}: IceCube is 
the largest neutrino detector in the world and has been operating from the South Pole for 
about 10 years. It has detected hundreds of astrophysical neutrinos with energies 
in some cases extending beyond 1 PeV ($10^{15}$ eV).} on September 22, 2017, in fact, was found to 
be in spatial coincidence with the known BL Lac TXS~0506+056 at $z=0.3365$ undergoing a 
period of enhanced $\gamma$-ray emission, first observed by the {\it Fermi} satellite and 
later by the MAGIC telescopes. The chance coincidence of the neutrino alert with the flaring $\gamma$-ray source is at the level of $3 - 3.5\sigma$~\citep{IceCube_2018a}.
Furthermore, an archival analysis revealed that during the September 2014 to March 2015 time period
TXS 0506+056 showed a prolonged outburst when it emitted $13\pm5$ neutrinos. 
A chance correlation of this type of neutrino outburst can be excluded at a confidence 
level of $3.5\sigma$~\citep{IceCube_2018b}. This association adds to a growing body of evidence, 
which links at least some blazars to IceCube neutrinos \citep[e.g.][and references therein]{Giommi_2020,Giommi_2021}. 

The importance of this multi-messenger connection can hardly be exaggerated, for various reasons.  
First, it has vital implications for the nature of $\gamma$-ray emission in blazars, which
is still debated, as two alternative, or possibly complementary, interpretations have been 
considered \citep[e.g.][and references therein]{Cerruti_2020}. The first scenario, so-called 
leptonic, explains $\gamma$-ray emission as inverse Compton scattering between the electrons 
in the jet and their own synchrotron emission (synchrotron self-Compton) or an external photon 
field (external inverse Compton) \citep{Maraschi_1992}. In the second scenario, the hadronic 
one, $\gamma$-rays are instead assumed to originate from high-energy protons either loosing energy 
through synchrotron emission \citep{Aharonian_2000} or through the photo-meson process 
\citep{Mannheim_1993}. The latter is the production of pions (or pi mesons: $\pi$) through 
proton-photon interactions, which generate $\pi^0, \pi^+$ and $\pi^-$. The neutral pion decades
into two $\gamma$-ray photons, while the charged pions end up producing electron/positron pairs
and (muon and electron) neutrinos (and anti-neutrinos). Photo-meson production has thus one 
fundamental property: neutrinos and $\gamma$-rays (of roughly similar energy and flux) are
produced together. Neutrino detection from a blazar is therefore the smoking gun that relativistic
protons, i.e. hadronic processes, are at work. Moreover, while $\gamma$-rays are absorbed by 
pair-production interactions with the extragalactic background light (EBL\footnote{The EBL is 
the sum of all the radiation produced by all galaxies in the Universe over its whole history.}) at $E \gtrsim
100$ GeV \citep[e.g.][and references therein]{Biteau_2020}, neutrinos can travel cosmological 
distances basically unaffected by matter and magnetic fields and are the only
``messengers'', which can provide information on the very high-energy physical processes
that generated them. Said differently, since the extragalactic photon sky is almost completely 
dark at the energies sampled by IceCube ($\gtrsim 60$ TeV), neutrinos are our only hope to probe
these energies.

\section{Circumnuclear matter on different physical scales}

The basic ingredient of standard Unification Models is an axisymmetric absorber, which prevents the direct view of the nuclear engine in edge-on line of sights (see Fig. \ref{fig:sketch}). The presence of this obscuring structure was inferred by the observation of broad emission lines and a big blue bump in the polarized light of Type 2 objects, thus unveiling their hidden Type 1 nucleus \citep[e.g.,][]{Antonucci1993}. Being the polarization very high and perpendicular to the radio axis, this is clear evidence that photons from the nuclear regions can only escape along the axis, and then get scattered into the line of sight. The equatorial direction is therefore blocked by some kind of dusty `toroidal' structure. 

In this scenario, all AGN are believed to have intrinsically the same nuclear engine, which emits an optical featureless continuum and broad emission lines, and an optically opaque torus, co-aligned with the radio axis. This torus should be located on parsec scales, in order to obscure the inner, dustless and high-density ($n_e\gtrsim10^8$ cm$^{-3}$) BLR, while leaving unaffected the outer, extended and low-density ($n_e\sim10^3-10^6$ cm$^{-3}$) NLR. In particular, the geometry of the torus, e.g. its opening angle, is postulated to be the same for all objects. However, it is now clear that this simplified scenario is not able to take into account all the complexities arisen in decades of multiwavelength observations. On the one hand, multiple absorbers, on different physical scales, must be now considered to picture a comprehensive view of obscuration in AGN. On the other hand, the X-ray perspective of circumnuclear matter in AGN has revolutionised our classic view of the standard ingredients of the Unification Models, i.e. the torus, the BLR and the NLR. In this respect, Compton-thick sources (absorbed by a column density [N$_\mathrm{H}$] of neutral gas larger than the inverse of the Thomson cross section, $\sigma_T^{-1}\simeq1.5\times10^{24}$ cm$^{-2}$) play a prominent role, since they allow for a clean view of the circumnuclear matter in the X-rays, not being overwhelmed by the nuclear emission.

\subsection{Within the sublimation radius}\label{sec:subl_r}

Dust sublimation is generally believed to set the outer boundary of the BLR \citep{Laor1993,Netzer1993a}. It occurs at a temperature around $1200-1800$ K, depending on the grain composition and size, so that a correspondent sublimation radius can be defined, which scales with the luminosity of the central source as $r_{sub} \propto L^{1/2}$ \citep[e.g.][]{Barvainis1987a}. Indeed, the size of the BLR is well known to scale with the square root of luminosity, with a tight relation which holds over several orders of magnitude \citep[e.g.][]{Bentz2013}. Most models picture the BLR as a disc wind, predicting its disappearance at very low luminosity or accretion rates \citep{Nicastro2000,Laor2003}. This was apparently confirmed by the existence of low accreting \textit{true} type 2 AGN, where the lack of broad optical lines is not associated with obscuration along the line of sight \citep[e.g.][for a comprehensive picture]{Bianchi2012}. However, a very broad H$\alpha$ line in the archetypal true type 2 NGC~3147 was revealed once a significant fraction of the host galaxy contamination was removed thanks to a small-aperture \textit{Hubble Space Telescope} (HST\footnote{\url{https://hubblesite.org/}}) aperture, extending the $R_{BLR}-L^{1/2}$ relation to very low accretion rates, and questioning the very existence of this class of objects \citep{Bianchi2019a}.

At odds with the standard Unification Model, X-ray absorption variability gives solid evidence in favour of the presence of a component of absorbing gas within the sublimation radius. In the last 20 years, it was shown that this is a common feature in AGN. First results were presented in \cite{Risaliti2002} with the analysis of a sample of nearby obscured AGN with multiple X-ray observations, showing almost ubiquitous  N$_\mathrm{H}$ variations in local Seyfert galaxies. This has a twofold physical implication: the X-ray absorber must be clumpy and located close enough to the central source to witness significant variability. It is interesting to stress that the presence of gas inside the dust sublimation radius necessarily produces a mismatch between the optical/near-IR reddening (extinction) and X-ray column density measured along the line of sight. Indeed, a decrease of the expected dust-to-gas ratio is a long-standing observational evidence in nearby Seyfert galaxies \citep[e.g.][]{Maccacaro1982,Maiolino2001}, and can be naturally explained by the effect of X-ray absorption by BLR clouds.
    
Constraints on the geometrical and physical properties of this gas were derived through several observational campaigns \citep[e.g.][]{Elvis2004,Risaliti2005a,Bianchi2009a}. In the archetypical case, NGC~1365, Compton-thin (N$_\mathrm{H}\sim10^{23}$~cm$^{-2}$) to thick (N$_H>10^{24}$~cm$^{-2}$) state transitions were observed on time scales as short as $\sim$10~hours. Assuming that the absorbing gas is moving with Keplerian velocity, these rapid events imply the presence of clouds at $\sim10^4$ r$_\mathrm{g}$\footnote{r$_\mathrm{g}=GM/c^2$ is the gravitational radius of an object of mass $M$, where $G$ is the Gravitational constant and $c$ the speed of light.} from the BH, rotating at velocities larger than $10^3$~km~s$^{-1}$. Estimates for the size and density of the clouds can be also derived, being of the order of 10$^{13}$~cm and 10$^{10}$-10$^{11}$~cm$^{-3}$, respectively. The overall scenario is therefore in agreement with the identification of this component of the X-ray absorber with the BLR clouds responsible for broad emission lines in the optical/UV.

A careful analysis of X-ray ``eclipses'' occurring when an absorbing cloud passes along the line of sight provided further information on its geometrical and physical properties. Again for NGC~1365, \cite{Maiolino2010} studied the time evolution of its covering factor and its column density, unveiling a ``cometary'' shape for the obscuring cloud, i.e. a high density head, followed by an elongated, lower density tail. These ``cometary'' clouds may represent fascinating insights for open issues in modeling the BLR in AGN. For example, the mass loss through the cometary tail could result in the destruction of the cloud within a few months \citep{Maiolino2010}. Therefore, there must be a process which continuously replenishes the gas clouds, likely from the accretion disk itself. Such a process could indeed be key for the dynamical equilibrium invoked for the long-standing problem of the stability of BLR clouds \citep[e.g.][for a classical review]{Osterbrock1986}. Another implication of the cometary structure of the BLR would be on the overall broad emission line profile. In this case, each cloud would contribute to the observed profile with a small, but not null, width, reducing dramatically the total number of clouds needed to reproduce the observed smooth profiles. This would be a more consistent solution with respect to current estimates with `point-like' BLR clouds, which require enormous cloud densities to achieve the high level of observed smoothness \citep[see e.g.][for NGC~4151]{Arav1998}, resulting in a nearly complete occupation of the available volume. Future, higher throughput X-ray observatories (e.g. \textit{Athena}\footnote{\url{https://www.the-athena-x-ray-observatory.eu/}}, \textit{eXTP}\footnote{\url{https://www.isdc.unige.ch/extp/}}) may fully disclosure the powerful potentiality of occultation events to measure the physical and geometrical properties of the BLR clouds.

\begin{figure}
\centering
\includegraphics[width=0.6\textwidth]{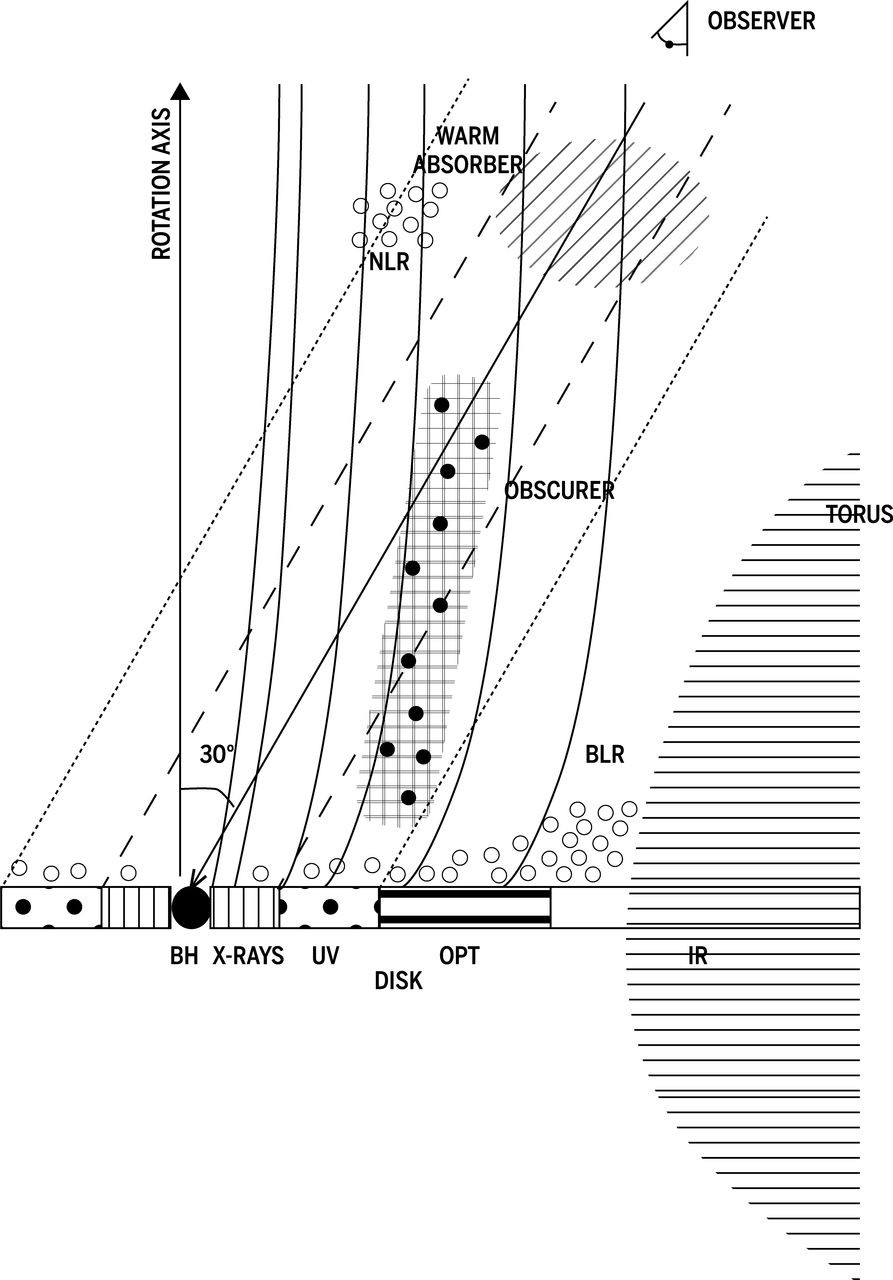}
\caption{A possible cartoon (not to scale) of the central region of an AGN. The accretion disk around the BH emits X-ray (via a corona), UV, optical (OPT), and IR continua at different scales. Outflows of gas are launched from the disk along the magnetic field lines (curved lines). Close to the inner BLR, an obscurer can be present, constituted by a cold dense component embedded in a more ionized gas. A dusty torus surrounds these regions, while the NLR and the warm absorber (WA) are farther out. Figure reproduced from \cite{Kaastra2014a}, Fig. 4, with permission.}
\label{fig:5548}       
\end{figure}

More recent observations have shown that such occultations can also occur in otherwise persistently unobscured sources. In particular, long-term X-ray monitorings discovered sudden spectral hardening events (i.e. a decrease of the soft X-ray flux with respect to the hard X-rays) in Type 1 objects, lasting for days up to several years \citep[e.g.][]{Markowitz2014,Mehdipour2017}. As first proposed for NGC~5548 \citep{Kaastra2014a}, these events are due to mildly ionized winds arising in the accretion disk, obscuring the nuclear source along the line of sight and possibly reaching beyond the BLR. This is depicted in Fig.~\ref{fig:5548}, where the location of this obscurer is shown with respect to the other basic components of the central region of an AGN. Interestingly, these obscuration events are often associated with UV broad absorption-lines \citep{Kriss2019}. They may therefore represent the shielding from the nuclear radiation required by
theoretical models of radiatively driven accretion-disk winds in Broad Absorption Line (BAL) QSOs \citep[e.g.][]{Proga2004a}. These obscurers are also found to be sometimes associated with highly ionized outflows \citep{Mehdipour2017}, outflowing at velocities that can be even relativistic as in the case of UFOs \cite{Nardini2015a}.

\subsection{The torus}

The standard Unified Model \citep{Antonucci1993,Urry_1995} postulated the presence of a circumnuclear dusty medium on parsec scales (see Fig.~\ref{fig:sketch}). This was soon confirmed by near-IR studies, with the evidence for very hot dust, close to the sublimation temperature, in the nuclei of Seyfert galaxies \citep{Storchi-Bergmann1992,Alonso-Herrero2001}. Subsequent extensive reverberation\footnote{Reverberation mapping is a  technique to determine the time-lag (hence the distance) between the continuum variability and that of reprocessed components from circumnuclear matter, such as dust re-emission from the torus or line emission from photoionized gas like the BLR.} observational campaigns \citep{Suganuma2006} confirmed the expected $L^{1/2}$ dependence of the sublimation radius, which lies indeed at sub-parsec scales for local Seyfert galaxies, and on parsec scales for higher-luminosity quasars. The near-IR observations also estimated a high covering factor for the circumnuclear dusty medium \citep[exceeding 0.8:][]{Maiolino2007a,Treister2008a}, thus in overall agreement with the observed  type 2/type 1 ratio \citep{Maiolino1995}.

The circumnuclear medium on parsec scales was effectively imaged by radio observations. Very Long Baseline Array (\textit{VLBA}\footnote{\url{https://science.nrao.edu/facilities/vlba}}) water maser\footnote{Water maser emission is due to a rotational transition in the H$_2$O molecule at 22 GHz, and traces high density ($10^7 < \mathrm{n_H} < 10^8$ cm$^{-3}$) and warm ($400 < T < 1000$ K) gas.} images at sub-parsec resolution revealed a rotating warped disk in NGC~1068 \citep{Greenhill1996}. Similar structures were obtained for other AGN \citep{Greenhill1997,Kondratko2008}. \textit{VLBA} images of the continuum radio emission of NGC~1068 further showed two symmetric radio free-free blobs at a radius of $\sim$0.3 pc with respect to the non-thermal nuclear emission, which was interpreted as the inner ionized edge of the obscuring torus \citep{Gallimore1997a}. Further details were provided by  \textit{ALMA}\footnote{\url{https://www.almaobservatory.org}} high-resolution submillimeter images, which revealed a compact, parsec-scale, gas component characterized by non-circular motions and enhanced turbulence \citep{Garcia-Burillo2016}. Interestingly, the torus appears to have a physical and dynamical connection
with  the  circumnuclear  disk  of  the  galaxy, a characteristic found in \textit{ALMA} images of other AGN \citep{Combes2019}.

Dust can be also imaged via mid-IR interferometry. This was achieved for the first time in 2004 for  NGC~1068 \citep{Jaffe2004}, and later refined by \cite{Raban2009}. They found that the dusty torus in this source has two components: a compact (0.5 pc thick), elongated hot ($T>800$ K) component, coincident with the nuclear water maser; and an extended (3-4 pc), but less elongated colder ($T\simeq300$ K) component. Similar results were found, with the same technique, for Circinus \citep{Tristram2007} and for a type 1 object, NGC~4151 \citep{Burtscher2009}. Indeed, when larger samples are taken into account, no significant differences are found between type 1 and 2 sources, with the size of the dusty emitter scaling as the square root of the luminosity, as previously stated \citep{Tristram2011,Kishimoto2011}. Combining radiative transfer modelling of the interferometric data with the study of the shape of the SED, there is now an emerging picture of a two-component structure: an equatorial thin disk and an extended feature along the polar direction probably originating from a dusty wind \citep[e.g.][]{Honig2017}. In general, the IR observational properties of the dusty pc-scale torus are better explained by a clumpy structure \citep{Nenkova2002,Elitzur2006a}, since a compact uniform torus can hardly reproduce the broad IR SED, which requires dust at multiple temperature \citep{Pier1992}. A complex structure for the central hot and warm dust in the torus of NGC~1068 was confirmed by a recent mid-IR spectro-interferometric MATISSE observation at VLTI\footnote{\url{https://www.eso.org/sci/facilities/paranal/telescopes/vlti.html}}, in an overall good agreement with the expectations from the original Unification Model \citep{GamezRosas2022}. 

Strong clues in favour of the ubiquitous presence of Compton-thick neutral material come also from the X-ray spectra of local Seyfert galaxies, which invariably show an iron K$\alpha$ line and a Compton reflection component \citep[e.g.][]{Perola2002,Bianchi2004,Bianchi2009}. Its covering factor is far from being universal and varies significantly from source to source \citep[e.g.][]{Marchesi2019}, with a clear trend of being smaller at increasing luminosity \citep{Bianchi2007a}. These components prevalently arise at least on parsec scales, as the standard torus envisaged in the Unification Models, since they generally lack any variability. Indeed, Compton-thick sources have X-ray spectra which are completely dominated by these `reflection' features, and they typically do not vary at all, even on long timescales. This is particularly evident in those `changing-look' AGN which can be interpreted as sources whose central engine shuts off, leaving a constant reflection component (including the Fe K$\alpha$ line) unaffected for years \citep[][]{Gilli2000,Matt2003}. 

However, it was recently found that the absorption in the archetypal Compton-thick source, NGC~1068, does vary, allowing the intrinsic emission of this buried AGN to pierce through, for the first time, during unveiling events \cite{Marinucci2016,Zaino2020}. The measured timescales for such events are compatible with obscuration at the inner part of the torus, close to the sublimation radius, and further suggest a clumpy structure.

\begin{figure}
\centering
\includegraphics[width=\textwidth]{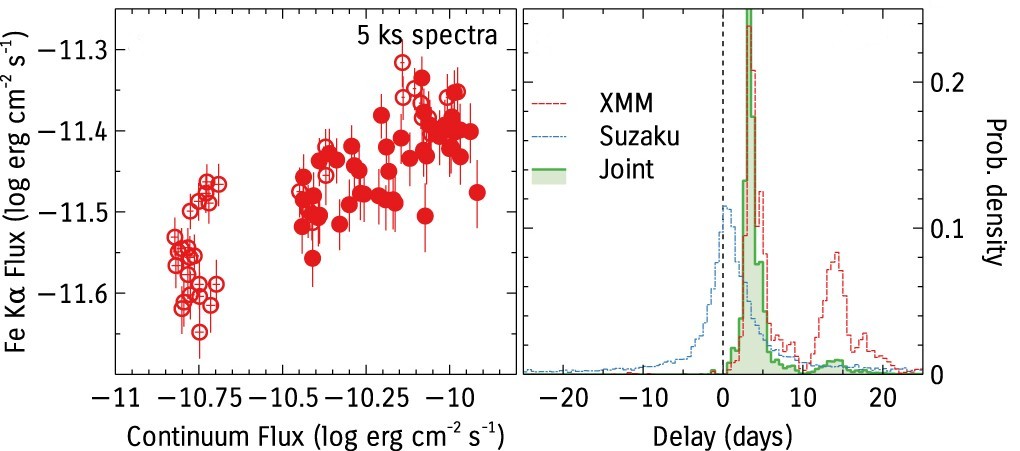}
\caption{Narrow Fe K$\alpha$ variability in NGC~4151. \textit{Left:} Variations of the emission line flux with respect to the observed 7-10 keV ionizing flux for 5 ks contiguous spectra. The observed strong correlation shows that the line responds to the continuum variability. \textit{Right:} Resulting probability density of the time delay between the narrow Fe K$\alpha$ line flux and the continuum flux. A peak at $3.3^{+1.8}_{-0.7}$ days suggests that the iron line is emitted from a region smaller than the optical BLR in this source. Figure adapted from \cite{Zoghbi2019}, Fig. 10, with permission.}
\label{fig:feka_4151}       
\end{figure}

Indeed, X-ray variability studies hold the potential to infer the geometry and distance of the torus. However, accurate X-ray reverberation analysis of the iron line and the Compton reflection component, with respect to the intrinsic variability of the primary continuum, is hampered by the limitations of current X-ray missions. To date, the only successful experiment was performed on NGC~4151, allowing \citet{Zoghbi2019} to derive an origin for the `narrow' core of the Fe K$\alpha$ in the inner BLR (Fig.~\ref{fig:feka_4151}). Future high-throughput X-ray telescopes, assisted by large field-of-view monitoring, would represent a breakthrough in this respect \citep[see e.g. this science case described in][for eXTP]{IntZand2018}.

Further clues come from the Fe K$\alpha$ line profile. A narrow core of this line is ubiquitous, at least in local Seyfert galaxies \citep{Bianchi2009}, and is typically unresolved (with upper limits of several thousands km/s for its FWHM), suggesting an origin at least in the BLR, or farther away. Although current X-ray satellites generally allows for separating the component originating in the accretion disk, being extremely broadened by strong gravity effects when present in the spectra \citep[see e.g.][]{DelaCallePerez2010a,Nandra2007a}, the unresolved core generally leads to inconclusive estimates on its location \citep[see e.g.][]{Nandra2006,Shu2011}. An exception is represented by the case of NGC~7213, which has been tested with simultaneous optical and X-ray observations. The iron line width in this source was resolved by the \textit{Chandra}\footnote{\url{https://chandra.harvard.edu/}} High-Energy Transmission Grating, and found to be in perfect agreement with that measured for the broad H$\alpha$ \citep{Bianchi2008a}. This suggests a common origin in the BLR, as further strengthened by the lack of a Compton-reflection component in this source, expected instead from a Compton-thick torus \citep{Bianchi2003,Bianchi2004}.
Micro-calorimeters, as those onboard \textit{XRISM}\footnote{\url{https://heasarc.gsfc.nasa.gov/docs/xrism/}} and \textit{Athena}, will revolutionize this kind of analysis, allowing us for the first time to deconvolve the different components of the iron line. For example, it will be possible to separate the contributions from the torus, the NLR and the BLR, giving an unprecedented map of the circumnuclear gas in AGN \citep{Cappi2013}.

\subsection{Beyond the torus up to the host galaxy}

\begin{figure}
\centering
\includegraphics[width=0.45\textwidth]{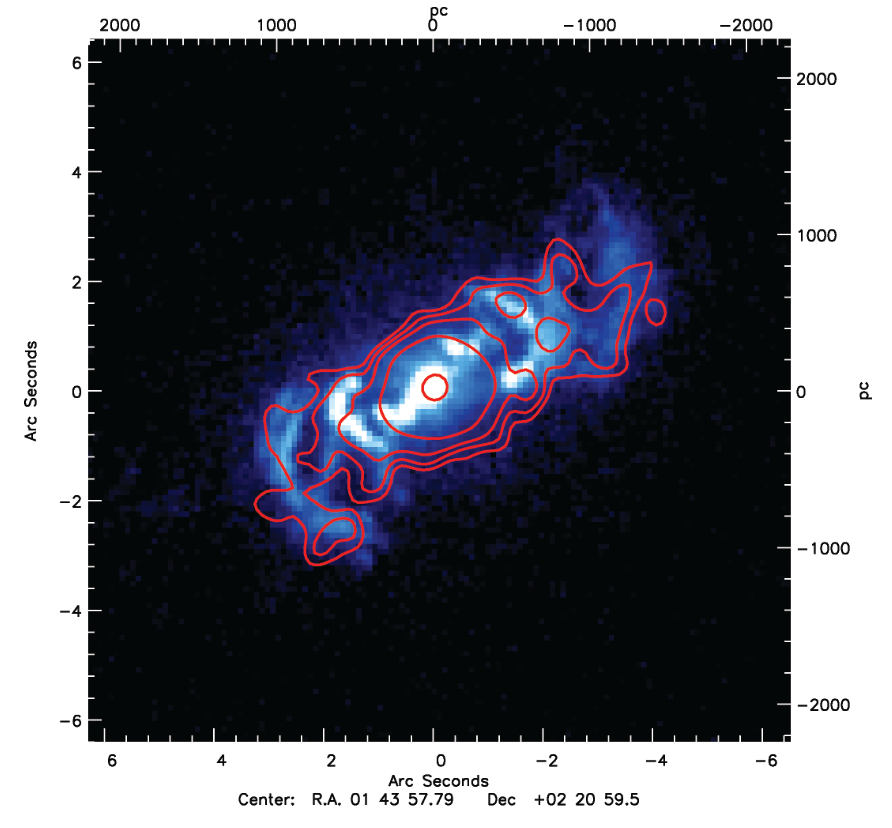}
\includegraphics[width=0.54\textwidth]{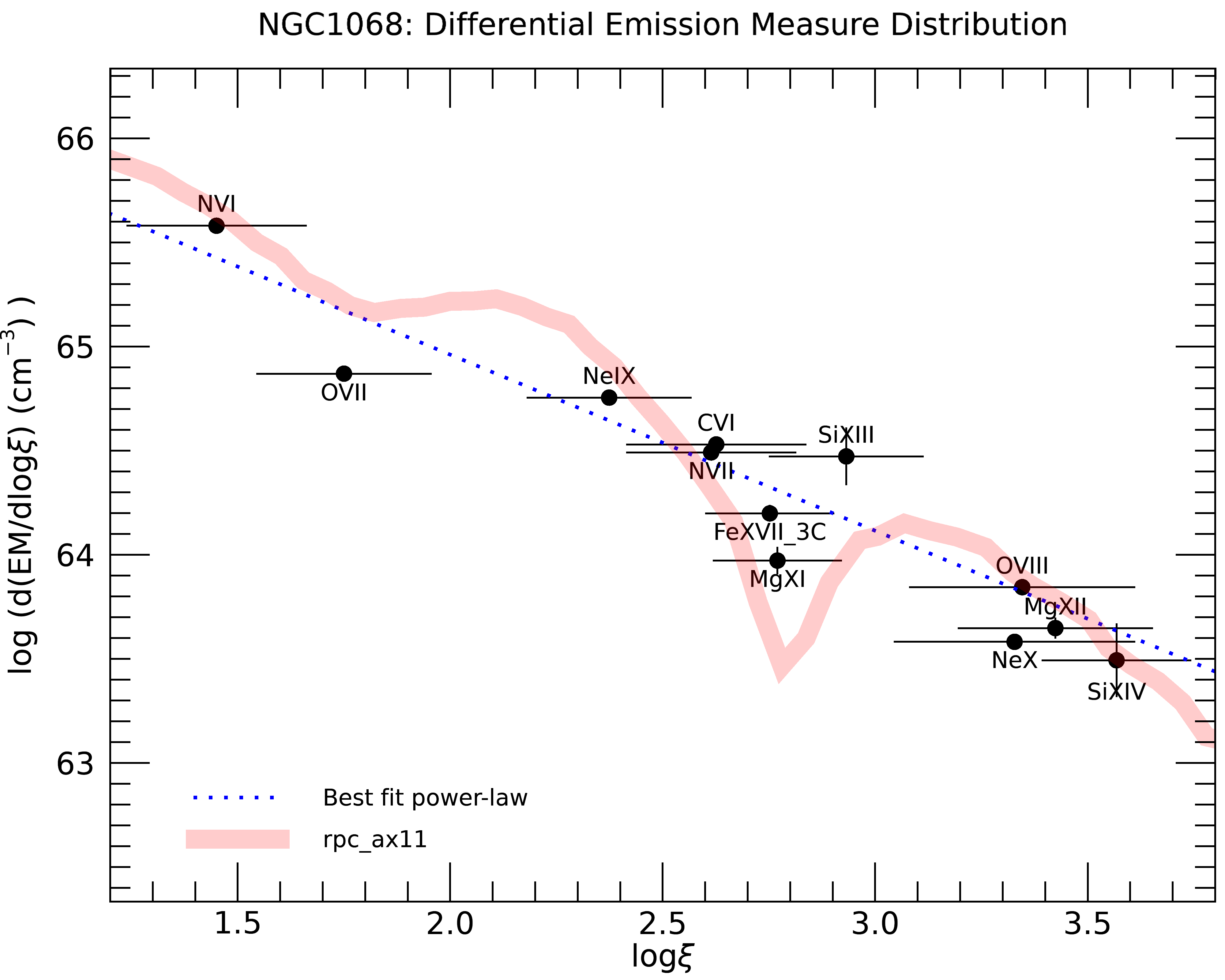}
\caption{\textit{Left:} \textit{Chandra} soft X-ray (0.2–2 keV) emission (contours) of the Seyfert 2 galaxy Mrk~573, superimposed on the \textit{HST} [O \textsc{iii}]$\lambda5007$\AA\ image, showing the morphological similarity. \textit{Right:} Observed Differential Emission Measure (DEM) for the soft X-ray emission of NGC~1068 (black circles), along with the best-fitting power law (dotted line). The DEM distribution for an Radiation Pressure Compressed gas (red shaded area) is in remarkable agreement with the data. Figures adapted from \cite{Bianchi2010a} and \cite{Bianchi2019}, Fig. 5 and 2, respectively, with permission.}
\label{fig:NLR}       
\end{figure}

Soon after the advent of high-resolution X-ray imaging with \textit{Chandra}, the soft X-ray emission of obscured Seyfert galaxy was found to be extended on at least hundreds of pc, with a remarkable morphological similarity with the optical NLR \citep[e.g.][]{Young2001a,Bianchi2006,Levenson2006,Fabbiano2018}. At the same time, high resolution X-ray spectroscopy, thanks to the gratings onboard XMM-\textit{Newton}\footnote{\url{https://www.cosmos.esa.int/web/xmm-newton}} and \textit{Chandra}, discovered that this emission is dominated by emission lines from a photoionized gas \citep[e.g.][]{Sako2000,Kinkhabwala2002,Guainazzi2007} (see left panel of Fig.~\ref{fig:NLR}). Optical and X-ray emission lines trace quite different ionization states of the gas, so their spatial overlap requires low and high-density components to co-exist in the NLR. This may be produced by classic thermal instabilities, whereas the same thermal pressure allows the high-density low-temperature gas to be confined by low-density high-temperature gas \citep{Krolik1981}. However, it was shown that this is instead an unavoidable consequence of the effect of radiation pressure, which compresses the gas in the illuminated cloud \citep[a mechanism known as Radiation Pressure Compression (RPC):][and references therein]{Stern2014a}. This scenario predicts a well-defined and universal density and ionization distribution for the NLR, which was nicely confirmed by the observed Differential Emission Measure (DEM) distribution (i.e. the contributions of each ionization zone to the total line flux) for the soft X-ray emission of a sample of obscured AGN \citep{Bianchi2019} (see right panel of Fig.~\ref{fig:NLR}).

Interestingly, an RPC wind well reproduces also the ionization distribution observed in the so-called ``warm absorbers'' \citep{Stern2014a,Goosmann2016a}. More than half of nearby Seyfert 1 galaxies show absorption from a photoionized gas, with a large range in ionization state, column density and outflow velocity \citep[e.g.][]{Blustin2005}. When a distance can be estimated for their location (i.e. via variability studies), typical warm absorbers are consistent with an origin in the NLR or the outer torus \citep[e.g.][]{Kaastra2012}. Indeed, a possible origin for this wind is via thermal evaporation of the torus \citep{Krolik2001}, although many other alternatives have been proposed \citep[e.g.][]{Contopoulos1994,Proga2000}.

High-resolution \textit{HST} images directly showed that dust lanes at distances of hundred of parsecs are commonly found in Seyfert galaxies \citep{Malkan1998}, partially obscuring the NLR itself. In some cases, the dust-lanes coincide with X-ray obscuration of the soft X-ray emission from the NLR \citep[e.g.][]{Bianchi2007}. Large distributions of gas at $\sim100$ pc from the AGN are also imaged in interferometric maps of the molecular gas, certainly contributing to the obscuration of the active nucleus \citep[e.g.][]{Schinnerer2000,Krips2011}. It was soon realized that optically selected AGN samples generally avoid edge-on systems, a clear sign that the host galaxy itself provides obscuration on 100 pc scales \citep{Maiolino1995,Lagos2011a}. Indeed, low column densities are often in agreement with the optical reddening associated with the host galaxy, and this must be taken into account in a more generalized Unification Model \citep[see e.g.][]{Maiolino1995,Matt2000,Bianchi2009a}. 
However, in general, the presence of these structures may not be physically responsible for the obscuration of the nucleus, but it appears to be correlated with Compton-thin X-ray obscuration \citep{Guainazzi2005}. Recent IR/optical dust maps confirm that dust lanes and filaments, similar to what is seen in the Milky Way, are present in both Type 1 and Type 2 objects, but only in the latter objects they cross the nucleus with enough optical depth to fully obscure it \citep{Prieto2021}.

In any case, a word of caution is needed for the amount of obscuring gas on large scales. Indeed, it was shown that the high covering factor needed to explain the observed Compton-thick sources significantly limits the presence of Compton-thick gas at distances of ~100 pc, in order not to exceed the dynamical mass in that region \citep{Risaliti1999}. An obvious consequence of this requirement is that most of the ubiquitous narrow neutral iron K$\alpha$ line and X-ray Compton reflection component should originate from a compact region. However, these components are seen extending up to $\simeq2$ kpc in NGC~1068 \citep{Young2001a}, and this seems to be common whenever high-quality \textit{Chandra} data are available \citep[e.g.][]{Marinucci2013a,Marinucci2017,Fabbiano2017,Jones2021a}.
Indeed, these extended components appear to be consistent with an origin in a Compton-thin gas, possibly X-ray dissociated dense molecular clouds of the host galaxy \cite[e.g.][and references therein]{Ma2020}.

\section{AGN demography and evolution in the X-ray and $\gamma$-ray bands}\label{sec:dem}

The study of the evolution of the number density and luminosity of the AGN population has been a topic of interest since the 1960's. Providing a complete census of AGN over cosmic time is a critical tool to understand how accretion onto SMBHs evolves and also the role these may have in the evolution of their own host galaxies. However, in order to achieve a complete AGN census it is crucial to be able to observe also obscured AGN, which may be efficiently done in the X-ray and $\gamma$-ray bands, as described in the following.

\subsection{X-ray band}

X-ray observations are one of the most reliable and complete methods for selecting AGN. They are reliable because the X-ray emission from other astrophysical processes is usually weak in comparison. They are complete due to the ability of X-ray photons, especially above a few keV, to penetrate through absorbing material. In fact, the deepest X-ray surveys performed with the {\it Chandra} and {\it XMM-Newton} satellites reached the highest sky density of detected AGN: 23,900 deg$^{-2}$ \cite[e.g.][see also Fig. \ref{fig:densities}]{Luo17}.

\begin{figure}
\centering
\includegraphics[width=0.49\textwidth]{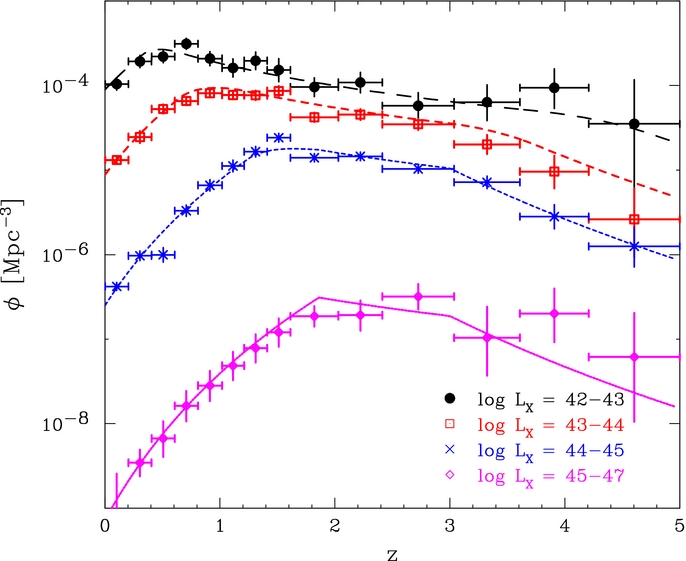}
\includegraphics[width=0.50\textwidth]{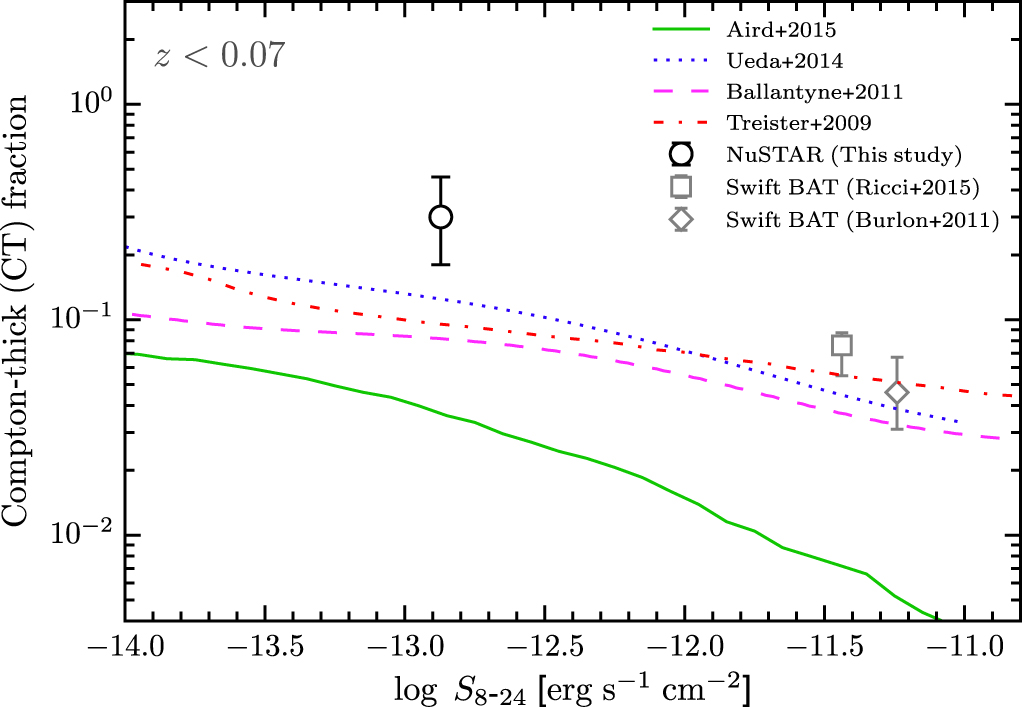}
\caption{\textit{Left:} Comoving number density of AGN plotted against redshift in different luminosity bins from \cite{ueda14}. The data points are the density of AGN as a function of redshift measured from the soft and hard X-ray samples used in \cite{ueda14}, while the curves are the best-fit model per luminosity bin. \textit{Right:}  Observed fraction of Compton-thick AGN as a function of the 8–24 keV flux limit, for $z< 0.07$ from \cite{lansbury17}; see also \cite{ananna19}. The data points are the measured Compton-thick fractions from {\it NuSTAR} and Swift/BAT observations, while the curves are model predictions. Figures reproduced from \cite{ueda14} and \cite{lansbury17}, Fig. 11 and Fig. 12, respectively, with permission.   }
\label{fig:XLF}       
\end{figure}

The study of the X-ray Luminosity Function (XLF) has been a primary tool since the discovery of AGN to trace how the density of such objects changes as a function of luminosity and redshift. Combining large area shallow X-ray surveys (e.g. {\it ROSAT}\footnote{\url{https://www.mpe.mpg.de/ROSAT}} and {\it ASCA} \footnote{\url{https://www.isas.jaxa.jp/en/missions/spacecraft/past/asca.html}}) with deep pencil beam ones (e.g. {\it Chandra} and {\it XMM-Newton}) several authors have studied how the shape and normalization of the XLF changes with luminosity and redshift. Many studies found that the overall XLF at z$\sim 0-3$ is well described by a so-called luminosity dependent density evolution (LDDE) model \citep[e.g.][]{miyaji00,ueda03,hasinger05,silverman08,ebrero09,ueda14}. In this 
model the shape of the XLF changes with redshift, with the faint-end slope flattening as redshift increases. This is a combination of 
the ``classic'' pure luminosity and pure density evolution models, in which only the luminosity and the number density 
of sources changes with redshift, respectively.  
On the other hand, \cite{aird15} modelled separately the XLF for unobscured and obscured AGN in the same redshifts range and found that a so-called flexible double power-law (FDPL) model is able to reproduce the evolution of the shape of the XLF. This schema models the XLF as a double power-law but allows for arbitrary evolution in the overall luminosity, density or shape of the function. A common feature of all these studies is the clear presence of a strong luminosity evolution: higher luminosity AGN peak, in terms of their space density, at higher redshifts than lower luminosity objects (Fig. \ref{fig:XLF} left panel). The underlying driver of such evolution could be a change in the mass distributions of accreting SMBHs: at earlier cosmic times SMBHs with higher masses are accreting, while the accretion moves to lower mass SMBHs going to lower redshifts. Another important finding is that the normalization of the XLF changes significantly  with redshift: it increases between z$\sim 0$ and z$\sim 1$ and decreases afterwards. The rise of the number density of accreating SMBHs could be a consequence of the increase in molecular gas content in galaxies at earlier cosmic epochs \citep[e.g.][for a recent review]{tacconi20} which could also be the main driver of the peak in star-formation density at z$\sim 1-3$ \citep[e.g.][]{madau14}. A particularly interesting, but challenging, task is to quantify the evolution of the accreting SMBH populations at high redshift. Using the deepest available Chandra surveys, \cite{vito18} confirmed the overall decline of the AGN population beyond redshifts z$\sim 3$ and, interestingly, seem to find a steeper decline for the low-luminosity population compared to the decline in galaxies density. Still, the overall density of low-luminosity AGN, especially at z$> 5$, is very poorly constrained. Future X-ray missions like {\it Athena} and {\it Lynx}\footnote{\url{https://www.lynxobservatory.com/}}, thanks to their larger collecting area, will be able to detect statistically significant samples of low-luminosity AGN at these high redshifts. Complementing these future X-ray surveys with spectroscopic follow-up campaigns with Extremely Large Telescopes (ELT\footnote{\url{https://elt.eso.org}}, TMT\footnote{\url{https://www.tmt.org}}, GMT\footnote{\url{https://www.gmto.org}}) or dedicated wide field spectroscopic facilities from the ground will provide redshifts, black hole masses and host galaxy properties, and  finally allow us to place stronger constraints on the accretion history of SMBH at these early epochs. 

As also discussed in the previous sections, obscuration both on nuclear and galactic scales can affect a significant fraction of the accretion history on SMBHs. Between the variety of methods used to identify obscured AGNs \citep[e.g.][for a recent review]{hickox18} X-ray surveys are one of the most complete. Since the impact of obscuration is a function of energy, with low-energy X-ray photons being more affected than high-energy ones, X-ray satellites sensitive to energies above $\approx 10$ keV (e.g. {\it Integral}\footnote{\url{https://sci.esa.int/web/integral/}}, {\it Swift-BAT}\footnote{\url{https://swift.gsfc.nasa.gov}}, {\it NuSTAR}\footnote{\url{https://www.nustar.caltech.edu}}) are more effective in detecting heavily obscured AGN than satellites mostly sensitive to softer energy bands (e.g. {\it Chandra}, {\it XMM-Newton}). This is attenuated moving to higher redshifts since also in the observed $<10$ keV band we would be sampling harder rest-frame X-ray photons. Several X-ray (but also at other wavelengths) studies have found a significant dependence of the obscured AGN fraction with luminosity \citep[e.g.][]{hasinger08,ueda14,merloni14,aird15,mateos17}. For example, \cite{merloni14}  reported that the fraction of X-ray obscured  (N$_{\rm H}>10^{22}$ cm$^{-2}$) AGN goes from $\approx 60\%$ at L$_{\rm X}=10^{43}$ erg s$^{-1}$ down to $\approx 20\%$ at L$_{\rm X}=10^{45}$ erg s$^{-1}$. A possible interpretation of this trend could be connected with the ability of the most powerful AGN to release an increasing radiation and momentum in the surrounding medium and effectively decrease the obscuration of the central source. A redshift dependence of the obscured fraction of AGN was originally more controversial, but it is now generally confirmed \citep[e.g.][]{lafranca05,hasinger08,aird15,ananna19}. The fraction of obscured objects increases going to earlier epochs, which is consistent, and plausibly correlated, with the increase of the gas fraction in galaxies \citep[e.g.][for a recent review]{tacconi20}. Finally, particularly challenging is the ability to quantify the fraction of Compton-thick AGN. The most stringent constraints, especially at low redshift, have been obtained with hard X-ray telescopes (e.g. {\it Swift/BAT}, {\it NuSTAR}) which returned a Compton-thick fraction around $\approx 30\%$  \citep[Fig. \ref{fig:XLF} right panel][]{burlon11,ricci15,lansbury17}. This fraction should be still considered a lower limit on the intrinsic population of Compton-thick. Further improvement will be provided from additional {\it NuSTAR} observations, as well as a future more sensitive hard X-ray mission (e.g. the High-Energy X-ray Probe study).

\subsection{$\gamma$-ray band}

AGN evolution in the $\gamma$-rays is driven by blazars, which totally dominate the 
$\gamma$-ray sky (Sect. \ref{sec:AGN_HE}). Blazar evolution has been studied in the 
{\it Fermi} band (50 MeV -- 1 TeV) but not at higher energies. This is due to the fact
that TeV-selected flux-limited samples are not available because of the lack of sensitive 
blind surveys due to the narrow field of views and limited sensitivities of IACTs (Sect. 
\ref{sec:AGN_HE}). Moreover,
observations of TeV blazars are often triggered by flaring states and therefore the measured 
TeV fluxes do not represent the typical condition of the sources. Therefore, our knowledge of 
blazar very high energy $\gamma$-ray emission is very biased and patchy. The Cherenkov Telescope Array\footnote{\url{https://www.cta-observatory.org/}} will adopt a systematic approach and will 
detect approximately 10 times more sources reaching also higher energies.  

The samples put together 
using {\it Fermi} Large Area Telescope (LAT) catalogues have been used to determine the population 
properties of blazars, i.e., their $\gamma$-ray number counts, luminosity function, and 
cosmological evolution. The main results of this work are the following: 1. the evolution 
of $\gamma$-ray selected blazars is very strong and has been modelled as primarily luminosity 
evolution (PLE), primarily density evolution (PDE), and  LDDE \cite{Ajello_2015}. As the names suggest, although in all models the sources experience 
both luminosity and (number) density evolution, the former and the latter are the dominant 
one in PLE and PDE, respectively; 2. the latest {\it Fermi}-LAT number count distribution of 
extragalactic sources (at $|b_{\rm II}| > 20^{\circ}$) show that PDE is the model that best 
describes the evolutionary properties of the blazar population and that blazars account 
for $\sim 50^{+10}_{-5}\%$ of the total extragalactic $\gamma$-ray background (EGB). This 
includes 
$\gamma$-ray emission from resolved and unresolved extragalactic sources, such as blazars, 
star-forming galaxies and radio galaxies, as well as radiation from truly diffuse processes, 
above 
100 MeV \cite{Marcotulli_2020}. Hard $\gamma$-ray spectrum sources, i.e., HBLs, are 
actually responsible for most of the EGB emission above 100 GeV \cite{Ajello_2015}; 
3. when considering only BL Lacs, however, a somewhat complex picture emerges, 
with luminous ($L_{\gamma} \sim 10^{47}$ erg s$^{-1}$) objects evolving as strongly as
FSRQs while less luminous BL Lacs evolve less until they evolve negatively,
i.e. their luminosity and/or density decreases with redshift, for $L_{\gamma} \lesssim 
10^{45.5}$ erg s$^{-1}$. Once the sample is subdivided into HBLs, IBLs, and LBLs, it 
turns out that the negative evolution is confined to HBLs, whose number density increases
for $z \lesssim 0.5$ \cite{Ajello_2012,Ajello_2014}. This confirms previous results derived
in other bands on the puzzling negative evolution of HBLs (e.g., \cite{Giommi_1999}). 
Given the much larger {\it Fermi}-LAT samples available now \cite{4LAC,4FGL-DR3}, it would be
useful to re-do these studies with a much larger statistics. 

\bibliographystyle{mnras}
\bibliography{Section12_chap1}

\begin{thebibliography}{}
\makeatletter
\relax
\def\mn@urlcharsother{\let\do\@makeother \do\$\do\&\do\#\do\^\do\_\do\%\do\~}
\def\mn@doi{\begingroup\mn@urlcharsother \@ifnextchar [ {\mn@doi@}
  {\mn@doi@[]}}
\def\mn@doi@[#1]#2{\def\@tempa{#1}\ifx\@tempa\@empty \href
  {http://dx.doi.org/#2} {doi:#2}\else \href {http://dx.doi.org/#2} {#1}\fi
  \endgroup}
\def\mn@eprint#1#2{\mn@eprint@#1:#2::\@nil}
\def\mn@eprint@arXiv#1{\href {http://arxiv.org/abs/#1} {{\tt arXiv:#1}}}
\def\mn@eprint@dblp#1{\href {http://dblp.uni-trier.de/rec/bibtex/#1.xml}
  {dblp:#1}}
\def\mn@eprint@#1:#2:#3:#4\@nil{\def\@tempa {#1}\def\@tempb {#2}\def\@tempc
  {#3}\ifx \@tempc \@empty \let \@tempc \@tempb \let \@tempb \@tempa \fi \ifx
  \@tempb \@empty \def\@tempb {arXiv}\fi \@ifundefined
  {mn@eprint@\@tempb}{\@tempb:\@tempc}{\expandafter \expandafter \csname
  mn@eprint@\@tempb\endcsname \expandafter{\@tempc}}}

\bibitem[\protect\citeauthoryear{{Abdo} et~al.,}{{Abdo}
  et~al.}{2010}]{Abdo_2010}
{Abdo} A.~A.,  et~al., 2010, \mn@doi [\apj] {10.1088/0004-637X/716/1/30}, \href
  {https://ui.adsabs.harvard.edu/abs/2010ApJ...716...30A} {716, 30}

\bibitem[\protect\citeauthoryear{{Abdollahi} et~al.,}{{Abdollahi}
  et~al.}{2020}]{4FGL}
{Abdollahi} S.,  et~al., 2020, \mn@doi [\apjs] {10.3847/1538-4365/ab6bcb},
  \href {https://ui.adsabs.harvard.edu/abs/2020ApJS..247...33A} {247, 33}

\bibitem[\protect\citeauthoryear{{Aharonian}}{{Aharonian}}{2000}]{Aharonian_2000}
{Aharonian} F.~A.,  2000, \mn@doi [\na] {10.1016/S1384-1076(00)00039-7}, \href
  {https://ui.adsabs.harvard.edu/abs/2000NewA....5..377A} {5, 377}

\bibitem[\protect\citeauthoryear{{Aird}, {Coil}, {Georgakakis}, {Nandra},
  {Barro}  \& {P{\'e}rez-Gonz{\'a}lez}}{{Aird} et~al.}{2015}]{aird15}
{Aird} J.,  {Coil} A.~L.,  {Georgakakis} A.,  {Nandra} K.,  {Barro} G.,
  {P{\'e}rez-Gonz{\'a}lez} P.~G.,  2015, \mn@doi [\mnras]
  {10.1093/mnras/stv1062}, \href
  {https://ui.adsabs.harvard.edu/abs/2015MNRAS.451.1892A} {451, 1892}

\bibitem[\protect\citeauthoryear{{Ajello} et~al.,}{{Ajello}
  et~al.}{2012}]{Ajello_2012}
{Ajello} M.,  et~al., 2012, \mn@doi [\apj] {10.1088/0004-637X/751/2/108}, \href
  {https://ui.adsabs.harvard.edu/abs/2012ApJ...751..108A} {751, 108}

\bibitem[\protect\citeauthoryear{{Ajello} et~al.,}{{Ajello}
  et~al.}{2014}]{Ajello_2014}
{Ajello} M.,  et~al., 2014, \mn@doi [\apj] {10.1088/0004-637X/780/1/73}, \href
  {https://ui.adsabs.harvard.edu/abs/2014ApJ...780...73A} {780, 73}

\bibitem[\protect\citeauthoryear{{Ajello} et~al.,}{{Ajello}
  et~al.}{2015}]{Ajello_2015}
{Ajello} M.,  et~al., 2015, \mn@doi [\apjl] {10.1088/2041-8205/800/2/L27},
  \href {https://ui.adsabs.harvard.edu/abs/2015ApJ...800L..27A} {800, L27}

\bibitem[\protect\citeauthoryear{{Ajello} et~al.,}{{Ajello}
  et~al.}{2020a}]{4LAC}
{Ajello} M.,  et~al., 2020a, \mn@doi [\apj] {10.3847/1538-4357/ab791e}, \href
  {https://ui.adsabs.harvard.edu/abs/2020ApJ...892..105A} {892, 105}

\bibitem[\protect\citeauthoryear{{Ajello}, {Di Mauro}, {Paliya}  \&
  {Garrappa}}{{Ajello} et~al.}{2020b}]{Ajello_2020}
{Ajello} M.,  {Di Mauro} M.,  {Paliya} V.~S.,   {Garrappa} S.,  2020b, \mn@doi
  [\apj] {10.3847/1538-4357/ab86a6}, \href
  {https://ui.adsabs.harvard.edu/abs/2020ApJ...894...88A} {894, 88}

\bibitem[\protect\citeauthoryear{{Ajello} et~al.,}{{Ajello}
  et~al.}{2021}]{Fermi_2021}
{Ajello} M.,  et~al., 2021, \mn@doi [\apj] {10.3847/1538-4357/ac1bb2}, \href
  {https://ui.adsabs.harvard.edu/abs/2021ApJ...921..144A} {921, 144}

\bibitem[\protect\citeauthoryear{Alonso-Herrero, Quillen, Simpson, Efstathiou
  \& Ward}{Alonso-Herrero et~al.}{2001}]{Alonso-Herrero2001}
Alonso-Herrero A.,  Quillen A.~C.,  Simpson C.,  Efstathiou A.,   Ward M.~J.,
  2001, \mn@doi [\aj] {10.1086/319410}, 121, 1369

\bibitem[\protect\citeauthoryear{{Ananna} et~al.,}{{Ananna}
  et~al.}{2019}]{ananna19}
{Ananna} T.~T.,  et~al., 2019, \mn@doi [\apj] {10.3847/1538-4357/aafb77}, \href
  {https://ui.adsabs.harvard.edu/abs/2019ApJ...871..240A} {871, 240}

\bibitem[\protect\citeauthoryear{Antonucci}{Antonucci}{1993}]{Antonucci1993}
Antonucci R.,  1993, \mn@doi [\araa] {10.1146/annurev.aa.31.090193.002353}, 31,
  473

\bibitem[\protect\citeauthoryear{Arav, Barlow, Laor, Sargent  \&
  Blandford}{Arav et~al.}{1998}]{Arav1998}
Arav N.,  Barlow T.~A.,  Laor A.,  Sargent W.~L.,   Blandford R.~D.,  1998,
  \mn@doi [\mnras] {10.1046/j.1365-8711.1998.297004990.x}, 297, 990

\bibitem[\protect\citeauthoryear{Barvainis}{Barvainis}{1987}]{Barvainis1987a}
Barvainis R.,  1987, \mn@doi [\apj] {10.1086/165571}, 320, 537

\bibitem[\protect\citeauthoryear{Bentz et~al.,}{Bentz et~al.}{2013}]{Bentz2013}
Bentz M.~C.,  et~al., 2013, \mn@doi [\apj] {10.1088/0004-637X/767/2/149}, 767,
  149

\bibitem[\protect\citeauthoryear{Bianchi, Matt, Balestra  \& Perola}{Bianchi
  et~al.}{2003}]{Bianchi2003}
Bianchi S.,  Matt G.,  Balestra I.,   Perola G.~C.,  2003, \mn@doi [\aap]
  {10.1051/0004-6361:20031054}, 407, L21

\bibitem[\protect\citeauthoryear{Bianchi, Matt, Balestra, Guainazzi  \&
  Perola}{Bianchi et~al.}{2004}]{Bianchi2004}
Bianchi S.,  Matt G.,  Balestra I.,  Guainazzi M.,   Perola G.~C.,  2004,
  \mn@doi [\aap] {10.1051/0004-6361:20047128}, 422, 65

\bibitem[\protect\citeauthoryear{Bianchi, Guainazzi  \& Chiaberge}{Bianchi
  et~al.}{2006}]{Bianchi2006}
Bianchi S.,  Guainazzi M.,   Chiaberge M.,  2006, \mn@doi [\aap]
  {10.1051/0004-6361:20054091}, 448, 499

\bibitem[\protect\citeauthoryear{Bianchi, Chiaberge, Piconcelli  \&
  Guainazzi}{Bianchi et~al.}{2007a}]{Bianchi2007}
Bianchi S.,  Chiaberge M.,  Piconcelli E.,   Guainazzi M.,  2007a, \mn@doi
  [\mnras] {10.1111/j.1365-2966.2006.11183.x}, 374, 697

\bibitem[\protect\citeauthoryear{Bianchi, Guainazzi, Matt  \& {Fonseca
  Bonilla}}{Bianchi et~al.}{2007b}]{Bianchi2007a}
Bianchi S.,  Guainazzi M.,  Matt G.,   {Fonseca Bonilla} N.,  2007b, \mn@doi
  [\aap] {10.1051/0004-6361:20077331}, 467, L19

\bibitem[\protect\citeauthoryear{Bianchi, {La Franca}, Matt, Guainazzi,
  {Jimenez Bail{\'{o}}n}, Longinotti, Nicastro  \& Pentericci}{Bianchi
  et~al.}{2008a}]{Bianchi2008a}
Bianchi S.,  {La Franca} F.,  Matt G.,  Guainazzi M.,  {Jimenez Bail{\'{o}}n}
  E.,  Longinotti A.~L.,  Nicastro F.,   Pentericci L.,  2008a, \mn@doi
  [\mnras] {10.1111/j.1745-3933.2008.00521.x}, 389, L52

\bibitem[\protect\citeauthoryear{Bianchi, Guainazzi, Matt, Bonilla  \&
  Ponti}{Bianchi et~al.}{2008b}]{Bianchi2009}
Bianchi S.,  Guainazzi M.,  Matt G.,  Bonilla N.~F.,   Ponti G.,  2008b,
  \mn@doi [\aap] {10.1051/0004-6361:200810620}, 495, 421

\bibitem[\protect\citeauthoryear{Bianchi, Piconcelli, Chiaberge, {Jimenez
  Bail{\'{o}}n}, Matt  \& Fiore}{Bianchi et~al.}{2009}]{Bianchi2009a}
Bianchi S.,  Piconcelli E.,  Chiaberge M.,  {Jimenez Bail{\'{o}}n} E.,  Matt
  G.,   Fiore F.,  2009, \mn@doi [\apj] {10.1088/0004-637X/695/1/781}, 695, 781

\bibitem[\protect\citeauthoryear{Bianchi, Chiaberge, Evans, Guainazzi, Baldi,
  Matt  \& Piconcelli}{Bianchi et~al.}{2010}]{Bianchi2010a}
Bianchi S.,  Chiaberge M.,  Evans D.~A.,  Guainazzi M.,  Baldi R.~D.,  Matt G.,
    Piconcelli E.,  2010, \mn@doi [\mnras] {10.1111/j.1365-2966.2010.16475.x},
  405, 553

\bibitem[\protect\citeauthoryear{Bianchi et~al.,}{Bianchi
  et~al.}{2012}]{Bianchi2012}
Bianchi S.,  et~al., 2012, \mn@doi [\mnras] {10.1111/j.1365-2966.2012.21959.x},
  426, 3225

\bibitem[\protect\citeauthoryear{Bianchi, Guainazzi, Laor, Stern  \&
  Behar}{Bianchi et~al.}{2019a}]{Bianchi2019}
Bianchi S.,  Guainazzi M.,  Laor A.,  Stern J.,   Behar E.,  2019a, \mn@doi
  [\mnras] {10.1093/mnras/stz430}, 485, 416

\bibitem[\protect\citeauthoryear{Bianchi et~al.,}{Bianchi
  et~al.}{2019b}]{Bianchi2019a}
Bianchi S.,  et~al., 2019b, \mn@doi [\mnras] {10.1093/mnrasl/slz080}, 488, L1

\bibitem[\protect\citeauthoryear{{Biteau} et~al.,}{{Biteau}
  et~al.}{2020}]{Biteau_2020}
{Biteau} J.,  et~al., 2020, \mn@doi [Nature Astronomy]
  {10.1038/s41550-019-0988-4}, \href
  {https://ui.adsabs.harvard.edu/abs/2020NatAs...4..124B} {4, 124}

\bibitem[\protect\citeauthoryear{Blustin, Page, Fuerst, Branduardi-Raymont  \&
  Ashton}{Blustin et~al.}{2005}]{Blustin2005}
Blustin A.~J.,  Page M.~J.,  Fuerst S.~V.,  Branduardi-Raymont G.,   Ashton
  C.~E.,  2005, \mn@doi [\aap] {10.1051/0004-6361:20041775}, 431, 111

\bibitem[\protect\citeauthoryear{{Brandt} \& {Alexander}}{{Brandt} \&
  {Alexander}}{2015}]{Brandt_2015}
{Brandt} W.~N.,  {Alexander} D.~M.,  2015, \mn@doi [\aapr]
  {10.1007/s00159-014-0081-z}, \href
  {https://ui.adsabs.harvard.edu/abs/2015A&ARv..23....1B} {23, 1}

\bibitem[\protect\citeauthoryear{{Burlon}, {Ajello}, {Greiner}, {Comastri},
  {Merloni}  \& {Gehrels}}{{Burlon} et~al.}{2011}]{burlon11}
{Burlon} D.,  {Ajello} M.,  {Greiner} J.,  {Comastri} A.,  {Merloni} A.,
  {Gehrels} N.,  2011, \mn@doi [\apj] {10.1088/0004-637X/728/1/58}, \href
  {https://ui.adsabs.harvard.edu/abs/2011ApJ...728...58B} {728, 58}

\bibitem[\protect\citeauthoryear{Burtscher, Jaffe, Raban, Meisenheimer,
  Tristram  \& R{\"{o}}ttgering}{Burtscher et~al.}{2009}]{Burtscher2009}
Burtscher L.,  Jaffe W.,  Raban D.,  Meisenheimer K.,  Tristram K.~R.,
  R{\"{o}}ttgering H.,  2009, \mn@doi [\apj] {10.1088/0004-637X/705/1/L53},
  705, 53

\bibitem[\protect\citeauthoryear{Cappi et~al.,}{Cappi et~al.}{2013}]{Cappi2013}
Cappi M.,  et~al., 2013

\bibitem[\protect\citeauthoryear{{Cerruti}}{{Cerruti}}{2020}]{Cerruti_2020}
{Cerruti} M.,  2020, \mn@doi [Galaxies] {10.3390/galaxies8040072}, \href
  {https://ui.adsabs.harvard.edu/abs/2020Galax...8...72C} {8, 72}

\bibitem[\protect\citeauthoryear{Combes et~al.,}{Combes
  et~al.}{2019}]{Combes2019}
Combes F.,  et~al., 2019, \mn@doi [\aap] {10.1051/0004-6361/201834560}, 623,
  A79

\bibitem[\protect\citeauthoryear{{Condon}, {Kellermann}, {Kimball},
  {Ivezi{\'c}}  \& {Perley}}{{Condon} et~al.}{2013}]{Condon_2013}
{Condon} J.~J.,  {Kellermann} K.~I.,  {Kimball} A.~E.,  {Ivezi{\'c}} {\v{Z}}.,
   {Perley} R.~A.,  2013, \mn@doi [\apj] {10.1088/0004-637X/768/1/37}, \href
  {https://ui.adsabs.harvard.edu/abs/2013ApJ...768...37C} {768, 37}

\bibitem[\protect\citeauthoryear{Contopoulos \& Lovelace}{Contopoulos \&
  Lovelace}{1994}]{Contopoulos1994}
Contopoulos J.,  Lovelace R. V.~E.,  1994, \mn@doi [\apj] {10.1086/174307},
  429, 139

\bibitem[\protect\citeauthoryear{{D'Onofrio}, {Marziani}  \&
  {Sulentic}}{{D'Onofrio} et~al.}{2012}]{Donofrio_2012}
{D'Onofrio} M.,  {Marziani} P.,   {Sulentic} J.~W.,  2012, {Fifty Years of
  Quasars}.
~ Vol. 386, \mn@doi{10.1007/978-3-642-27564-7, }

\bibitem[\protect\citeauthoryear{{Delvecchio} et~al.,}{{Delvecchio}
  et~al.}{2017}]{Delvecchio_2017}
{Delvecchio} I.,  et~al., 2017, \mn@doi [\aap] {10.1051/0004-6361/201629367},
  \href {https://ui.adsabs.harvard.edu/abs/2017A&A...602A...3D} {602, A3}

\bibitem[\protect\citeauthoryear{{Ebrero} et~al.,}{{Ebrero}
  et~al.}{2009}]{ebrero09}
{Ebrero} J.,  et~al., 2009, \mn@doi [\aap] {10.1051/0004-6361:200810919}, \href
  {https://ui.adsabs.harvard.edu/abs/2009A&A...493...55E} {493, 55}

\bibitem[\protect\citeauthoryear{Elitzur \& Shlosman}{Elitzur \&
  Shlosman}{2006}]{Elitzur2006a}
Elitzur M.,  Shlosman I.,  2006, \mn@doi [\apj] {10.1086/508158}, 648, L101

\bibitem[\protect\citeauthoryear{{Elvis} et~al.,}{{Elvis}
  et~al.}{1994}]{Elvis_1994}
{Elvis} M.,  et~al., 1994, \mn@doi [\apjs] {10.1086/192093}, \href
  {https://ui.adsabs.harvard.edu/abs/1994ApJS...95....1E} {95, 1}

\bibitem[\protect\citeauthoryear{Elvis, Risaliti, Nicastro, Miller, Fiore  \&
  Puccetti}{Elvis et~al.}{2004}]{Elvis2004}
Elvis M.,  Risaliti G.,  Nicastro F.,  Miller J.~M.,  Fiore F.,   Puccetti S.,
  2004, \mn@doi [\apj] {10.1086/424380}, 615, L25

\bibitem[\protect\citeauthoryear{Fabbiano, Elvis, Paggi, Karovska, Maksym,
  Raymond, Risaliti  \& Wang}{Fabbiano et~al.}{2017}]{Fabbiano2017}
Fabbiano G.,  Elvis M.,  Paggi A.,  Karovska M.,  Maksym W.~P.,  Raymond J.,
  Risaliti G.,   Wang J.,  2017, \mn@doi [\apj] {10.3847/2041-8213/aa7551},
  842, L4

\bibitem[\protect\citeauthoryear{Fabbiano, Paggi, Karovska, Elvis, Maksym  \&
  Wang}{Fabbiano et~al.}{2018}]{Fabbiano2018}
Fabbiano G.,  Paggi A.,  Karovska M.,  Elvis M.,  Maksym W.~P.,   Wang J.,
  2018, eprint arXiv:1808.06985

\bibitem[\protect\citeauthoryear{{Fanaroff} \& {Riley}}{{Fanaroff} \&
  {Riley}}{1974}]{Fanaroff_1974}
{Fanaroff} B.~L.,  {Riley} J.~M.,  1974, \mn@doi [\mnras]
  {10.1093/mnras/167.1.31P}, \href
  {https://ui.adsabs.harvard.edu/abs/1974MNRAS.167P..31F} {167, 31P}

\bibitem[\protect\citeauthoryear{{Fermi-LAT collaboration} et~al.,}{{Fermi-LAT
  collaboration} et~al.}{2022}]{4FGL-DR3}
{Fermi-LAT collaboration} et~al., 2022, arXiv e-prints, \href
  {https://ui.adsabs.harvard.edu/abs/2022arXiv220111184F} {p. arXiv:2201.11184}

\bibitem[\protect\citeauthoryear{Gallimore, Baum  \& O'Dea}{Gallimore
  et~al.}{1997}]{Gallimore1997a}
Gallimore J.~F.,  Baum S.~A.,   O'Dea C.~P.,  1997, \mn@doi [Nature]
  {10.1038/42201}, 388, 852

\bibitem[\protect\citeauthoryear{{G{\'a}mez Rosas} et~al.,}{{G{\'a}mez Rosas}
  et~al.}{2022}]{GamezRosas2022}
{G{\'a}mez Rosas} V.,  et~al., 2022, \mn@doi [\nat]
  {10.1038/s41586-021-04311-7}, \href
  {https://ui.adsabs.harvard.edu/abs/2022Natur.602..403G} {602, 403}

\bibitem[\protect\citeauthoryear{Garc{\'{i}}a-Burillo
  et~al.,}{Garc{\'{i}}a-Burillo et~al.}{2016}]{Garcia-Burillo2016}
Garc{\'{i}}a-Burillo S.,  et~al., 2016, \mn@doi [\apj]
  {10.3847/2041-8205/823/1/l12}, 823, L12

\bibitem[\protect\citeauthoryear{Gilli, Maiolino, Marconi, Risaliti, Dadina,
  Weaver  \& Colbert}{Gilli et~al.}{2000}]{Gilli2000}
Gilli R.,  Maiolino R.,  Marconi A.,  Risaliti G.,  Dadina M.,  Weaver K.~A.,
  Colbert E. J.~M.,  2000, \aap, 355, 485

\bibitem[\protect\citeauthoryear{{Giommi} \& {Padovani}}{{Giommi} \&
  {Padovani}}{2021}]{Giommi_2021}
{Giommi} P.,  {Padovani} P.,  2021, \mn@doi [Universe]
  {10.3390/universe7120492}, \href
  {https://ui.adsabs.harvard.edu/abs/2021Univ....7..492G} {7, 492}

\bibitem[\protect\citeauthoryear{{Giommi}, {Menna}  \& {Padovani}}{{Giommi}
  et~al.}{1999}]{Giommi_1999}
{Giommi} P.,  {Menna} M.~T.,   {Padovani} P.,  1999, \mn@doi [\mnras]
  {10.1046/j.1365-8711.1999.02942.x}, \href
  {https://ui.adsabs.harvard.edu/abs/1999MNRAS.310..465G} {310, 465}

\bibitem[\protect\citeauthoryear{{Giommi}, {Glauch}, {Padovani}, {Resconi},
  {Turcati}  \& {Chang}}{{Giommi} et~al.}{2020}]{Giommi_2020}
{Giommi} P.,  {Glauch} T.,  {Padovani} P.,  {Resconi} E.,  {Turcati} A.,
  {Chang} Y.~L.,  2020, \mn@doi [\mnras] {10.1093/mnras/staa2082}, \href
  {https://ui.adsabs.harvard.edu/abs/2020MNRAS.497..865G} {497, 865}

\bibitem[\protect\citeauthoryear{Goosmann, Holczer, Mouchet, Dumont, Behar,
  Godet, Goncalves  \& Kaspi}{Goosmann et~al.}{2016}]{Goosmann2016a}
Goosmann R.~W.,  Holczer T.,  Mouchet M.,  Dumont A.~M.,  Behar E.,  Godet O.,
  Goncalves A.~C.,   Kaspi S.,  2016, \mn@doi [\aap]
  {10.1051/0004-6361/201425199}, 589, 1

\bibitem[\protect\citeauthoryear{Greenhill, Gwinn, Antonucci  \&
  Barvainis}{Greenhill et~al.}{1996}]{Greenhill1996}
Greenhill L.~J.,  Gwinn C.~R.,  Antonucci R.,   Barvainis R.,  1996, \mn@doi
  [\apj] {10.1086/310346}, 472, L21

\bibitem[\protect\citeauthoryear{Greenhill, Moran  \& Herrnstein}{Greenhill
  et~al.}{1997}]{Greenhill1997}
Greenhill L.~J.,  Moran J.~M.,   Herrnstein J.~R.,  1997, \mn@doi [\apj]
  {10.1086/310643}, 481, L23

\bibitem[\protect\citeauthoryear{Guainazzi \& Bianchi}{Guainazzi \&
  Bianchi}{2007}]{Guainazzi2007}
Guainazzi M.,  Bianchi S.,  2007, \mn@doi [\mnras]
  {10.1111/j.1365-2966.2006.11229.x}, 374, 1290

\bibitem[\protect\citeauthoryear{Guainazzi, Matt  \& Perola}{Guainazzi
  et~al.}{2005}]{Guainazzi2005}
Guainazzi M.,  Matt G.,   Perola G.~C.,  2005, \mn@doi [Astron. Astrophys. Vol.
  444, Issue 1, December II 2005, pp.119-132] {10.1051/0004-6361:20053643},
  444, 119

\bibitem[\protect\citeauthoryear{{Hasinger}}{{Hasinger}}{2008}]{hasinger08}
{Hasinger} G.,  2008, \mn@doi [\aap] {10.1051/0004-6361:200809839}, \href
  {https://ui.adsabs.harvard.edu/abs/2008A&A...490..905H} {490, 905}

\bibitem[\protect\citeauthoryear{{Hasinger}, {Miyaji}  \& {Schmidt}}{{Hasinger}
  et~al.}{2005}]{hasinger05}
{Hasinger} G.,  {Miyaji} T.,   {Schmidt} M.,  2005, \mn@doi [\aap]
  {10.1051/0004-6361:20042134}, \href
  {https://ui.adsabs.harvard.edu/abs/2005A&A...441..417H} {441, 417}

\bibitem[\protect\citeauthoryear{{Heckman} \& {Best}}{{Heckman} \&
  {Best}}{2014}]{Heckman_2014}
{Heckman} T.~M.,  {Best} P.~N.,  2014, \mn@doi [\araa]
  {10.1146/annurev-astro-081913-035722}, \href
  {https://ui.adsabs.harvard.edu/abs/2014ARA&A..52..589H} {52, 589}

\bibitem[\protect\citeauthoryear{{Hickox} \& {Alexander}}{{Hickox} \&
  {Alexander}}{2018}]{hickox18}
{Hickox} R.~C.,  {Alexander} D.~M.,  2018, \mn@doi [\araa]
  {10.1146/annurev-astro-081817-051803}, \href
  {https://ui.adsabs.harvard.edu/abs/2018ARA&A..56..625H} {56, 625}

\bibitem[\protect\citeauthoryear{H{\"{o}}nig \& Kishimoto}{H{\"{o}}nig \&
  Kishimoto}{2017}]{Honig2017}
H{\"{o}}nig S.~F.,  Kishimoto M.,  2017, \mn@doi [\apj]
  {10.3847/2041-8213/aa6838}, 838, L20

\bibitem[\protect\citeauthoryear{{IceCube Collaboration}}{{IceCube
  Collaboration}}{2018}]{IceCube_2018b}
{IceCube Collaboration} 2018, \mn@doi [Science] {10.1126/science.aat2890},
  \href {https://ui.adsabs.harvard.edu/abs/2018Sci...361..147I} {361, 147}

\bibitem[\protect\citeauthoryear{{IceCube Collaboration et al.}}{{IceCube
  Collaboration et al.}}{2018}]{IceCube_2018a}
{IceCube Collaboration et al.} 2018, \mn@doi [Science]
  {10.1126/science.aat1378}, \href
  {https://ui.adsabs.harvard.edu/abs/2018Sci...361.1378I} {361, eaat1378}

\bibitem[\protect\citeauthoryear{Jaffe et~al.,}{Jaffe et~al.}{2004}]{Jaffe2004}
Jaffe W.,  et~al., 2004, \mn@doi [Nature] {10.1038/nature02531}, 429, 47

\bibitem[\protect\citeauthoryear{Jones et~al.,}{Jones
  et~al.}{2021}]{Jones2021a}
Jones M.~L.,  et~al., 2021, \mn@doi [\apj] {10.3847/1538-4357/abe128}, 910, 19

\bibitem[\protect\citeauthoryear{Kaastra et~al.,}{Kaastra
  et~al.}{2012}]{Kaastra2012}
Kaastra J.~S.,  et~al., 2012, \mn@doi [\aap] {10.1051/0004-6361/201118161},
  539, 17

\bibitem[\protect\citeauthoryear{Kaastra et~al.,}{Kaastra
  et~al.}{2014}]{Kaastra2014a}
Kaastra J.~S.,  et~al., 2014, \mn@doi [Science (80-. ).]
  {10.1126/science.1253787}, pp science.1253787--

\bibitem[\protect\citeauthoryear{{Kellermann}}{{Kellermann}}{2015}]{Kellermann_2015}
{Kellermann} K.~I.,  2015, in {Massaro} F.,  {Cheung} C.~C.,  {Lopez} E.,
  {Siemiginowska} A.,  eds, ~ Vol. 313, Extragalactic Jets from Every Angle. pp
  190--195 (\mn@eprint {arXiv} {1412.7867}), \mn@doi{10.1017/S1743921315002185}

\bibitem[\protect\citeauthoryear{{King} \& {Pounds}}{{King} \&
  {Pounds}}{2015}]{King_2015}
{King} A.,  {Pounds} K.,  2015, \mn@doi [\araa]
  {10.1146/annurev-astro-082214-122316}, \href
  {https://ui.adsabs.harvard.edu/abs/2015ARA&A..53..115K} {53, 115}

\bibitem[\protect\citeauthoryear{Kinkhabwala et~al.,}{Kinkhabwala
  et~al.}{2002}]{Kinkhabwala2002}
Kinkhabwala A.,  et~al., 2002, \mn@doi [\apj] {10.1086/341482}, 575, 732

\bibitem[\protect\citeauthoryear{Kishimoto, H{\"{o}}nig, Antonucci, Barvainis,
  Kotani, Tristram, Weigelt  \& Levin}{Kishimoto et~al.}{2011}]{Kishimoto2011}
Kishimoto M.,  H{\"{o}}nig S.~F.,  Antonucci R.,  Barvainis R.,  Kotani T.,
  Tristram K.~R.,  Weigelt G.,   Levin K.,  2011, \mn@doi [\aap]
  {10.1051/0004-6361/201016054}, 527, 121

\bibitem[\protect\citeauthoryear{Kondratko, Greenhill  \& Moran}{Kondratko
  et~al.}{2008}]{Kondratko2008}
Kondratko P.~T.,  Greenhill L.~J.,   Moran J.~M.,  2008, \mn@doi [\apj]
  {10.1086/586879}, 678, 87

\bibitem[\protect\citeauthoryear{Krips et~al.,}{Krips et~al.}{2011}]{Krips2011}
Krips M.,  et~al., 2011, \mn@doi [\apj] {10.1088/0004-637X/736/1/37}, 736, 37

\bibitem[\protect\citeauthoryear{Kriss et~al.,}{Kriss et~al.}{2019}]{Kriss2019}
Kriss G.~A.,  et~al., 2019, \mn@doi [\aap] {10.1051/0004-6361/201834326}, 621,
  A12

\bibitem[\protect\citeauthoryear{Krolik \& Kriss}{Krolik \&
  Kriss}{2001}]{Krolik2001}
Krolik J.~H.,  Kriss G.~A.,  2001, \mn@doi [\apj] {10.1086/323442}, 561, 684

\bibitem[\protect\citeauthoryear{Krolik, McKee  \& Tarter}{Krolik
  et~al.}{1981}]{Krolik1981}
Krolik J.~H.,  McKee C.~F.,   Tarter C.~B.,  1981, \mn@doi [\apj]
  {10.1086/159303}, 249, 422

\bibitem[\protect\citeauthoryear{{La Franca} et~al.,}{{La Franca}
  et~al.}{2005}]{lafranca05}
{La Franca} F.,  et~al., 2005, \mn@doi [\apj] {10.1086/497586}, \href
  {https://ui.adsabs.harvard.edu/abs/2005ApJ...635..864L} {635, 864}

\bibitem[\protect\citeauthoryear{Lagos, Padilla, Strauss, Cora  \& Hao}{Lagos
  et~al.}{2011}]{Lagos2011a}
Lagos C. d.~P.,  Padilla N.~D.,  Strauss M.~A.,  Cora S.~A.,   Hao L.,  2011,
  \mn@doi [\mnras] {10.1111/j.1365-2966.2011.18531.x}, 414, 2148

\bibitem[\protect\citeauthoryear{Lamastra, Menci, Fiore, Antonelli,
  Colafrancesco, Guetta  \& Stamerra}{Lamastra et~al.}{2017}]{Lamastra_2017}
Lamastra A.,  Menci N.,  Fiore F.,  Antonelli L.~A.,  Colafrancesco S.,  Guetta
  D.,   Stamerra A.,  2017, \mn@doi [Astron. Astrophys.]
  {10.1051/0004-6361/201731452}, 607, A18

\bibitem[\protect\citeauthoryear{{Lansbury} et~al.,}{{Lansbury}
  et~al.}{2017}]{lansbury17}
{Lansbury} G.~B.,  et~al., 2017, \mn@doi [\apj] {10.3847/1538-4357/aa8176},
  \href {https://ui.adsabs.harvard.edu/abs/2017ApJ...846...20L} {846, 20}

\bibitem[\protect\citeauthoryear{Laor}{Laor}{2003}]{Laor2003}
Laor A.,  2003, \mn@doi [\apj] {10.1086/375008}, 590, 86

\bibitem[\protect\citeauthoryear{Laor \& Draine}{Laor \&
  Draine}{1993}]{Laor1993}
Laor A.,  Draine B.~T.,  1993, \mn@doi [\apj] {10.1086/172149}, 402, 441

\bibitem[\protect\citeauthoryear{Levenson, Heckman, Krolik, Weaver  \&
  {$\dot{Z}$}ycki}{Levenson et~al.}{2006}]{Levenson2006}
Levenson N.~A.,  Heckman T.~M.,  Krolik J.~H.,  Weaver K.~A.,   {$\dot{Z}$}ycki
  P.~T.,  2006, \mn@doi [\apj] {10.1086/505735}, 648, 111

\bibitem[\protect\citeauthoryear{{Liu}, {Murase}, {Inoue}, {Ge}  \&
  {Wang}}{{Liu} et~al.}{2018}]{Liu_2018}
{Liu} R.-Y.,  {Murase} K.,  {Inoue} S.,  {Ge} C.,   {Wang} X.-Y.,  2018,
  \mn@doi [\apj] {10.3847/1538-4357/aaba74}, \href
  {https://ui.adsabs.harvard.edu/abs/2018ApJ...858....9L} {858, 9}

\bibitem[\protect\citeauthoryear{{Luo} et~al.,}{{Luo} et~al.}{2017}]{Luo17}
{Luo} B.,  et~al., 2017, \mn@doi [\apjs] {10.3847/1538-4365/228/1/2}, \href
  {https://ui.adsabs.harvard.edu/abs/2017ApJS..228....2L} {228, 2}

\bibitem[\protect\citeauthoryear{Ma, Elvis, Fabbiano, Balokovi{\'{c}}, Maksym,
  Jones  \& Risaliti}{Ma et~al.}{2020}]{Ma2020}
Ma J.,  Elvis M.,  Fabbiano G.,  Balokovi{\'{c}} M.,  Maksym W.~P.,  Jones
  M.~L.,   Risaliti G.,  2020, \mn@doi [\apj] {10.3847/1538-4357/abacbe}, 900,
  164

\bibitem[\protect\citeauthoryear{Maccacaro, Perola  \& Elvis}{Maccacaro
  et~al.}{1982}]{Maccacaro1982}
Maccacaro T.,  Perola G.~C.,   Elvis M.,  1982, \mn@doi [\apj]
  {10.1086/159961}, 257, 47

\bibitem[\protect\citeauthoryear{{Madau} \& {Dickinson}}{{Madau} \&
  {Dickinson}}{2014}]{madau14}
{Madau} P.,  {Dickinson} M.,  2014, \mn@doi [\araa]
  {10.1146/annurev-astro-081811-125615}, \href
  {https://ui.adsabs.harvard.edu/abs/2014ARA&A..52..415M} {52, 415}

\bibitem[\protect\citeauthoryear{Maiolino \& Rieke}{Maiolino \&
  Rieke}{1995}]{Maiolino1995}
Maiolino R.,  Rieke G.~H.,  1995, \mn@doi [\apj] {10.1086/176468}, 454, 95

\bibitem[\protect\citeauthoryear{Maiolino, Marconi, Salvati, Risaliti,
  Severgnini, Oliva, {La Franca}  \& Vanzi}{Maiolino
  et~al.}{2001}]{Maiolino2001}
Maiolino R.,  Marconi A.,  Salvati M.,  Risaliti G.,  Severgnini P.,  Oliva E.,
   {La Franca} F.,   Vanzi L.,  2001, \mn@doi [\aap]
  {10.1051/0004-6361:20000177}, 365, 28

\bibitem[\protect\citeauthoryear{Maiolino, Shemmer, Imanishi, Netzer, Oliva,
  Lutz  \& Sturm}{Maiolino et~al.}{2007}]{Maiolino2007a}
Maiolino R.,  Shemmer O.,  Imanishi M.,  Netzer H.,  Oliva E.,  Lutz D.,
  Sturm E.,  2007, \mn@doi [\aap] {10.1051/0004-6361:20077252}, 468, 979

\bibitem[\protect\citeauthoryear{Maiolino et~al.,}{Maiolino
  et~al.}{2010}]{Maiolino2010}
Maiolino R.,  et~al., 2010, \mn@doi [\aap] {10.1051/0004-6361/200913985}, 517,
  47

\bibitem[\protect\citeauthoryear{Malkan, Gorjian  \& Tam}{Malkan
  et~al.}{1998}]{Malkan1998}
Malkan M.~A.,  Gorjian V.,   Tam R.,  1998, \mn@doi [\apjs] {10.1086/313110},
  117, 25

\bibitem[\protect\citeauthoryear{{Mannheim}}{{Mannheim}}{1993}]{Mannheim_1993}
{Mannheim} K.,  1993, \aap, \href
  {https://ui.adsabs.harvard.edu/abs/1993A&A...269...67M} {269, 67}

\bibitem[\protect\citeauthoryear{{Maraschi}, {Ghisellini}  \&
  {Celotti}}{{Maraschi} et~al.}{1992}]{Maraschi_1992}
{Maraschi} L.,  {Ghisellini} G.,   {Celotti} A.,  1992, \mn@doi [\apjl]
  {10.1086/186531}, \href
  {https://ui.adsabs.harvard.edu/abs/1992ApJ...397L...5M} {397, L5}

\bibitem[\protect\citeauthoryear{{Marchesi} et~al.,}{{Marchesi}
  et~al.}{2019}]{Marchesi2019}
{Marchesi} S.,  et~al., 2019, \mn@doi [\apj] {10.3847/1538-4357/aafbeb}, \href
  {https://ui.adsabs.harvard.edu/abs/2019ApJ...872....8M} {872, 8}

\bibitem[\protect\citeauthoryear{{Marcotulli}, {Di Mauro}  \&
  {Ajello}}{{Marcotulli} et~al.}{2020}]{Marcotulli_2020}
{Marcotulli} L.,  {Di Mauro} M.,   {Ajello} M.,  2020, \mn@doi [\apj]
  {10.3847/1538-4357/ab8cbd}, \href
  {https://ui.adsabs.harvard.edu/abs/2020ApJ...896....6M} {896, 6}

\bibitem[\protect\citeauthoryear{Marinucci, Miniutti, Bianchi, Matt  \&
  Risaliti}{Marinucci et~al.}{2013}]{Marinucci2013a}
Marinucci A.,  Miniutti G.,  Bianchi S.,  Matt G.,   Risaliti G.,  2013,
  \mn@doi [\mnras] {10.1093/mnras/stt1759}, 436, 2500

\bibitem[\protect\citeauthoryear{Marinucci et~al.,}{Marinucci
  et~al.}{2016}]{Marinucci2016}
Marinucci A.,  et~al., 2016, \mn@doi [\mnras] {10.1093/mnrasl/slv178}, 456, L94

\bibitem[\protect\citeauthoryear{Marinucci, Bianchi, Fabbiano, Matt, Risaliti,
  Nardini  \& Wang}{Marinucci et~al.}{2017}]{Marinucci2017}
Marinucci A.,  Bianchi S.,  Fabbiano G.,  Matt G.,  Risaliti G.,  Nardini E.,
  Wang J.,  2017, \mn@doi [\mnras] {10.1093/mnras/stx1551}, 470, 4039

\bibitem[\protect\citeauthoryear{Markowitz, Krumpe  \& Nikutta}{Markowitz
  et~al.}{2014}]{Markowitz2014}
Markowitz A.~G.,  Krumpe M.,   Nikutta R.,  2014, \mn@doi [\mnras]
  {10.1093/mnras/stt2492}, 439, 1403

\bibitem[\protect\citeauthoryear{{Mateos} et~al.,}{{Mateos}
  et~al.}{2017}]{mateos17}
{Mateos} S.,  et~al., 2017, \mn@doi [\apjl] {10.3847/2041-8213/aa7268}, \href
  {https://ui.adsabs.harvard.edu/abs/2017ApJ...841L..18M} {841, L18}

\bibitem[\protect\citeauthoryear{Matt}{Matt}{2000}]{Matt2000}
Matt G.,  2000, \aap, 355

\bibitem[\protect\citeauthoryear{Matt, Bianchi, Guainazzi, Brandt, Fabian,
  Iwasawa  \& Perola}{Matt et~al.}{2003}]{Matt2003}
Matt G.,  Bianchi S.,  Guainazzi M.,  Brandt W.~N.,  Fabian A.~C.,  Iwasawa K.,
    Perola G.~C.,  2003, \mn@doi [\aap] {10.1051/0004-6361:20021817}, 399, 519

\bibitem[\protect\citeauthoryear{Mehdipour et~al.,}{Mehdipour
  et~al.}{2017}]{Mehdipour2017}
Mehdipour M.,  et~al., 2017, \mn@doi [\aap] {10.1051/0004-6361/201731175}, 607,
  A28

\bibitem[\protect\citeauthoryear{{Merloni} \& {Heinz}}{{Merloni} \&
  {Heinz}}{2013}]{Merloni_2013}
{Merloni} A.,  {Heinz} S.,  2013, {Evolution of Active Galactic Nuclei}.
p.~503, \mn@doi{10.1007/978-94-007-5609-0\_11}

\bibitem[\protect\citeauthoryear{{Merloni} et~al.,}{{Merloni}
  et~al.}{2014}]{merloni14}
{Merloni} A.,  et~al., 2014, \mn@doi [\mnras] {10.1093/mnras/stt2149}, \href
  {https://ui.adsabs.harvard.edu/abs/2014MNRAS.437.3550M} {437, 3550}

\bibitem[\protect\citeauthoryear{{Miyaji}, {Hasinger}  \& {Schmidt}}{{Miyaji}
  et~al.}{2000}]{miyaji00}
{Miyaji} T.,  {Hasinger} G.,   {Schmidt} M.,  2000, \aap, \href
  {https://ui.adsabs.harvard.edu/abs/2000A&A...353...25M} {353, 25}

\bibitem[\protect\citeauthoryear{Nandra}{Nandra}{2006}]{Nandra2006}
Nandra K.,  2006, {On the origin of the iron K$\alpha$ line cores in active
  galactic nuclei}, \mn@doi{10.1111/j.1745-3933.2006.00158.x}

\bibitem[\protect\citeauthoryear{Nandra, O'Neill, George  \& Reeves}{Nandra
  et~al.}{2007}]{Nandra2007a}
Nandra K.,  O'Neill P.~M.,  George I.~M.,   Reeves J.~N.,  2007, {An XMM-Newton
  survey of broad iron lines in Seyfert galaxies},
  \mn@doi{10.1111/j.1365-2966.2007.12331.x}

\bibitem[\protect\citeauthoryear{Nardini et~al.,}{Nardini
  et~al.}{2015}]{Nardini2015a}
Nardini E.,  et~al., 2015, \mn@doi [Science (80-. ).]
  {10.1126/science.1259202}, 347, 860

\bibitem[\protect\citeauthoryear{Nenkova, Ivezi{\'{c}}  \& Elitzur}{Nenkova
  et~al.}{2002}]{Nenkova2002}
Nenkova M.,  Ivezi{\'{c}} {\v{Z}}.,   Elitzur M.,  2002, \mn@doi [\apj]
  {10.1086/340857}, 570, L9

\bibitem[\protect\citeauthoryear{{Netzer}}{{Netzer}}{2015}]{Netzer_2015}
{Netzer} H.,  2015, \mn@doi [\araa] {10.1146/annurev-astro-082214-122302},
  \href {https://ui.adsabs.harvard.edu/abs/2015ARA&A..53..365N} {53, 365}

\bibitem[\protect\citeauthoryear{Netzer \& Laor}{Netzer \&
  Laor}{1993}]{Netzer1993a}
Netzer H.,  Laor A.,  1993, \mn@doi [\apj] {10.1086/186741}, 404, L51

\bibitem[\protect\citeauthoryear{Nicastro}{Nicastro}{2002}]{Nicastro2000}
Nicastro F.,  2002, \mn@doi [\apj] {10.1086/312491}, 530, L65

\bibitem[\protect\citeauthoryear{{Osterbrock} \& {Mathews}}{{Osterbrock} \&
  {Mathews}}{1986}]{Osterbrock1986}
{Osterbrock} D.~E.,  {Mathews} W.~G.,  1986, \mn@doi [\araa]
  {10.1146/annurev.aa.24.090186.001131}, \href
  {https://ui.adsabs.harvard.edu/abs/1986ARA&A..24..171O} {24, 171}

\bibitem[\protect\citeauthoryear{{Padovani}}{{Padovani}}{2016}]{Padovani_2016}
{Padovani} P.,  2016, \mn@doi [\aapr] {10.1007/s00159-016-0098-6}, \href
  {https://ui.adsabs.harvard.edu/abs/2016A&ARv..24...13P} {24, 13}

\bibitem[\protect\citeauthoryear{{Padovani}}{{Padovani}}{2017}]{Padovani_2017a}
{Padovani} P.,  2017, \mn@doi [Nature Astronomy] {10.1038/s41550-017-0194},
  \href {https://ui.adsabs.harvard.edu/abs/2017NatAs...1E.194P} {1, 0194}

\bibitem[\protect\citeauthoryear{{Padovani} \& {Giommi}}{{Padovani} \&
  {Giommi}}{1995}]{Padovani_1995}
{Padovani} P.,  {Giommi} P.,  1995, \mn@doi [\apj] {10.1086/175631}, \href
  {https://ui.adsabs.harvard.edu/abs/1995ApJ...444..567P} {444, 567}

\bibitem[\protect\citeauthoryear{{Padovani} et~al.,}{{Padovani}
  et~al.}{2017}]{Padovani_2017b}
{Padovani} P.,  et~al., 2017, \mn@doi [The Astronomy and Astrophysics Review]
  {10.1007/s00159-017-0102-9}, \href
  {https://ui.adsabs.harvard.edu/abs/2017A&ARv..25....2P} {25, 2}

\bibitem[\protect\citeauthoryear{Perola, Matt, Cappi, Fiore, Guainazzi,
  Maraschi, Petrucci  \& Piro}{Perola et~al.}{2002}]{Perola2002}
Perola G.~C.,  Matt G.,  Cappi M.,  Fiore F.,  Guainazzi M.,  Maraschi L.,
  Petrucci P.~O.,   Piro L.,  2002, \mn@doi [\aap]
  {10.1051/0004-6361:20020658}, 389, 802

\bibitem[\protect\citeauthoryear{Pier \& Krolik}{Pier \&
  Krolik}{1992}]{Pier1992}
Pier E.~A.,  Krolik J.~H.,  1992, \mn@doi [\apj] {10.1086/186597}, 399, L23

\bibitem[\protect\citeauthoryear{Prieto, Nadolny, Fern{\'{a}}ndez-Ontiveros  \&
  Mezcua}{Prieto et~al.}{2021}]{Prieto2021}
Prieto A.,  Nadolny J.,  Fern{\'{a}}ndez-Ontiveros J.~A.,   Mezcua M.,  2021,
  MNRAS, 000, 1

\bibitem[\protect\citeauthoryear{Proga \& Kallman}{Proga \&
  Kallman}{2004}]{Proga2004a}
Proga D.,  Kallman T.~R.,  2004, \mn@doi [\apj] {10.1086/425117}, 616, 688

\bibitem[\protect\citeauthoryear{Proga, Stone  \& Kallman}{Proga
  et~al.}{2000}]{Proga2000}
Proga D.,  Stone J.~M.,   Kallman T.~R.,  2000, \mn@doi [\apj]
  {10.1086/317154}, 543, 686

\bibitem[\protect\citeauthoryear{Raban, Jaffe, R{\"{o}}ttgering, Meisenheimer
  \& Tristram}{Raban et~al.}{2009}]{Raban2009}
Raban D.,  Jaffe W.,  R{\"{o}}ttgering H.,  Meisenheimer K.,   Tristram K.~R.,
  2009, \mn@doi [\mnras] {10.1111/j.1365-2966.2009.14439.x}, 394, 1325

\bibitem[\protect\citeauthoryear{{Radcliffe}, {Barthel}, {Garrett}, {Beswick},
  {Thomson}  \& {Muxlow}}{{Radcliffe} et~al.}{2021}]{Radcliffe_2021}
{Radcliffe} J.~F.,  {Barthel} P.~D.,  {Garrett} M.~A.,  {Beswick} R.~J.,
  {Thomson} A.~P.,   {Muxlow} T.~W.~B.,  2021, \mn@doi [\aap]
  {10.1051/0004-6361/202140791}, \href
  {https://ui.adsabs.harvard.edu/abs/2021A&A...649L...9R} {649, L9}

\bibitem[\protect\citeauthoryear{{Ramos Almeida} \& {Ricci}}{{Ramos Almeida} \&
  {Ricci}}{2017}]{RamosAlmeida2017}
{Ramos Almeida} C.,  {Ricci} C.,  2017, \mn@doi [Nature Astronomy]
  {10.1038/s41550-017-0232-z}, \href
  {https://ui.adsabs.harvard.edu/abs/2017NatAs...1..679R} {1, 679}

\bibitem[\protect\citeauthoryear{{Ricci}, {Ueda}, {Koss}, {Trakhtenbrot},
  {Bauer}  \& {Gandhi}}{{Ricci} et~al.}{2015}]{ricci15}
{Ricci} C.,  {Ueda} Y.,  {Koss} M.~J.,  {Trakhtenbrot} B.,  {Bauer} F.~E.,
  {Gandhi} P.,  2015, \mn@doi [\apjl] {10.1088/2041-8205/815/1/L13}, \href
  {https://ui.adsabs.harvard.edu/abs/2015ApJ...815L..13R} {815, L13}

\bibitem[\protect\citeauthoryear{{Richards} et~al.,}{{Richards}
  et~al.}{2006}]{Richards_2006}
{Richards} G.~T.,  et~al., 2006, \mn@doi [\apjs] {10.1086/506525}, \href
  {https://ui.adsabs.harvard.edu/abs/2006ApJS..166..470R} {166, 470}

\bibitem[\protect\citeauthoryear{Risaliti, Maiolino  \& Salvati}{Risaliti
  et~al.}{1999}]{Risaliti1999}
Risaliti G.,  Maiolino R.,   Salvati M.,  1999, \mn@doi [\apj]
  {10.1086/307623}, 522, 157

\bibitem[\protect\citeauthoryear{Risaliti, Elvis  \& Nicastro}{Risaliti
  et~al.}{2002}]{Risaliti2002}
Risaliti G.,  Elvis M.,   Nicastro F.,  2002, \mn@doi [\apj] {10.1086/324146},
  571, 234

\bibitem[\protect\citeauthoryear{Risaliti, Elvis, Fabbiano, Baldi  \&
  Zezas}{Risaliti et~al.}{2005}]{Risaliti2005a}
Risaliti G.,  Elvis M.,  Fabbiano G.,  Baldi A.,   Zezas A.,  2005, \mn@doi
  [\apj] {10.1086/430252}, 623, L93

\bibitem[\protect\citeauthoryear{Sako, Kahn, Paerels  \& Liedahl}{Sako
  et~al.}{2000}]{Sako2000}
Sako M.,  Kahn S.~M.,  Paerels F.,   Liedahl D.~A.,  2000, \mn@doi [\apj]
  {10.1086/317053}, 542, 684

\bibitem[\protect\citeauthoryear{{Sarzi} et~al.,}{{Sarzi}
  et~al.}{2010}]{Sarzi2010}
{Sarzi} M.,  et~al., 2010, \mn@doi [\mnras] {10.1111/j.1365-2966.2009.16039.x},
  \href {https://ui.adsabs.harvard.edu/abs/2010MNRAS.402.2187S} {402, 2187}

\bibitem[\protect\citeauthoryear{Schinnerer, Eckart, Tacconi, Genzel  \&
  Downes}{Schinnerer et~al.}{2000}]{Schinnerer2000}
Schinnerer E.,  Eckart A.,  Tacconi L.~J.,  Genzel R.,   Downes D.,  2000,
  \mn@doi [\apj] {10.1086/308702}, 533, 850

\bibitem[\protect\citeauthoryear{{Schmidt}}{{Schmidt}}{1963}]{Schmidt_1963}
{Schmidt} M.,  1963, \mn@doi [\nat] {10.1038/1971040a0}, \href
  {https://ui.adsabs.harvard.edu/abs/1963Natur.197.1040S} {197, 1040}

\bibitem[\protect\citeauthoryear{Shu, Yaqoob  \& Wang}{Shu
  et~al.}{2011}]{Shu2011}
Shu X.~W.,  Yaqoob T.,   Wang J.~X.,  2011, \mn@doi [\apj]
  {10.1088/0004-637X/738/2/147}, 738, 147

\bibitem[\protect\citeauthoryear{{Silverman} et~al.,}{{Silverman}
  et~al.}{2008}]{silverman08}
{Silverman} J.~D.,  et~al., 2008, \mn@doi [\apj] {10.1086/529572}, \href
  {https://ui.adsabs.harvard.edu/abs/2008ApJ...679..118S} {679, 118}

\bibitem[\protect\citeauthoryear{Stern, Behar, Laor, Baskin  \& Holczer}{Stern
  et~al.}{2014}]{Stern2014a}
Stern J.,  Behar E.,  Laor A.,  Baskin A.,   Holczer T.,  2014, \mn@doi
  [\mnras] {10.1093/mnras/stu1960}, 445, 3011

\bibitem[\protect\citeauthoryear{Storchi-Bergmann, Mulchaey  \&
  Wilson}{Storchi-Bergmann et~al.}{1992}]{Storchi-Bergmann1992}
Storchi-Bergmann T.,  Mulchaey J.~S.,   Wilson A.~S.,  1992, \mn@doi [\apj]
  {10.1086/186491}, 395, L73

\bibitem[\protect\citeauthoryear{Suganuma et~al.,}{Suganuma
  et~al.}{2006}]{Suganuma2006}
Suganuma M.,  et~al., 2006, \mn@doi [\apj] {10.1086/499326}, 639, 46

\bibitem[\protect\citeauthoryear{{Tacconi}, {Genzel}  \& {Sternberg}}{{Tacconi}
  et~al.}{2020}]{tacconi20}
{Tacconi} L.~J.,  {Genzel} R.,   {Sternberg} A.,  2020, \mn@doi [\araa]
  {10.1146/annurev-astro-082812-141034}, \href
  {https://ui.adsabs.harvard.edu/abs/2020ARA&A..58..157T} {58, 157}

\bibitem[\protect\citeauthoryear{{Tananbaum} et~al.,}{{Tananbaum}
  et~al.}{1979}]{tananbaum79}
{Tananbaum} H.,  et~al., 1979, \mn@doi [\apjl] {10.1086/183100}, \href
  {https://ui.adsabs.harvard.edu/abs/1979ApJ...234L...9T} {234, L9}

\bibitem[\protect\citeauthoryear{Treister, Krolik  \& Dullemond}{Treister
  et~al.}{2008}]{Treister2008a}
Treister E.,  Krolik J.~H.,   Dullemond C.,  2008, \mn@doi [\apj]
  {10.1086/586698}, 679, 140

\bibitem[\protect\citeauthoryear{Tristram \& Schartmann}{Tristram \&
  Schartmann}{2011}]{Tristram2011}
Tristram K.~R.,  Schartmann M.,  2011, \mn@doi [\aap]
  {10.1051/0004-6361/201116867}, 531, 99

\bibitem[\protect\citeauthoryear{Tristram et~al.,}{Tristram
  et~al.}{2007}]{Tristram2007}
Tristram K.~R.,  et~al., 2007, \mn@doi [\aap] {10.1051/0004-6361:20078369},
  474, 837

\bibitem[\protect\citeauthoryear{{Ueda}, {Akiyama}, {Ohta}  \& {Miyaji}}{{Ueda}
  et~al.}{2003}]{ueda03}
{Ueda} Y.,  {Akiyama} M.,  {Ohta} K.,   {Miyaji} T.,  2003, \mn@doi [\apj]
  {10.1086/378940}, \href
  {https://ui.adsabs.harvard.edu/abs/2003ApJ...598..886U} {598, 886}

\bibitem[\protect\citeauthoryear{{Ueda}, {Akiyama}, {Hasinger}, {Miyaji}  \&
  {Watson}}{{Ueda} et~al.}{2014}]{ueda14}
{Ueda} Y.,  {Akiyama} M.,  {Hasinger} G.,  {Miyaji} T.,   {Watson} M.~G.,
  2014, \mn@doi [\apj] {10.1088/0004-637X/786/2/104}, \href
  {https://ui.adsabs.harvard.edu/abs/2014ApJ...786..104U} {786, 104}

\bibitem[\protect\citeauthoryear{{Ulrich}, {Maraschi}  \& {Urry}}{{Ulrich}
  et~al.}{1997}]{Ulrich_1997}
{Ulrich} M.-H.,  {Maraschi} L.,   {Urry} C.~M.,  1997, \mn@doi [\araa]
  {10.1146/annurev.astro.35.1.445}, \href
  {https://ui.adsabs.harvard.edu/abs/1997ARA&A..35..445U} {35, 445}

\bibitem[\protect\citeauthoryear{{Urry} \& {Padovani}}{{Urry} \&
  {Padovani}}{1995}]{Urry_1995}
{Urry} C.~M.,  {Padovani} P.,  1995, \mn@doi [\pasp] {10.1086/133630}, \href
  {https://ui.adsabs.harvard.edu/abs/1995PASP..107..803U} {107, 803}

\bibitem[\protect\citeauthoryear{{Vito} et~al.,}{{Vito} et~al.}{2018}]{vito18}
{Vito} F.,  et~al., 2018, \mn@doi [\mnras] {10.1093/mnras/stx2486}, \href
  {https://ui.adsabs.harvard.edu/abs/2018MNRAS.473.2378V} {473, 2378}

\bibitem[\protect\citeauthoryear{{Wang} et~al.,}{{Wang}
  et~al.}{2021}]{Wang_2021}
{Wang} F.,  et~al., 2021, \mn@doi [\apjl] {10.3847/2041-8213/abd8c6}, \href
  {https://ui.adsabs.harvard.edu/abs/2021ApJ...907L...1W} {907, L1}

\bibitem[\protect\citeauthoryear{{Younes}, {Porquet}, {Sabra}, {Reeves}  \&
  {Grosso}}{{Younes} et~al.}{2012}]{Younes_2012}
{Younes} G.,  {Porquet} D.,  {Sabra} B.,  {Reeves} J.~N.,   {Grosso} N.,  2012,
  \mn@doi [\aap] {10.1051/0004-6361/201118299}, \href
  {https://ui.adsabs.harvard.edu/abs/2012A&A...539A.104Y} {539, A104}

\bibitem[\protect\citeauthoryear{Young, Wilson  \& Shopbell}{Young
  et~al.}{2001}]{Young2001a}
Young A.~J.,  Wilson A.~S.,   Shopbell P.~L.,  2001, \mn@doi [\apj]
  {10.1086/321561}, 556, 6

\bibitem[\protect\citeauthoryear{{Yuan} \& {Narayan}}{{Yuan} \&
  {Narayan}}{2014}]{Feng_2014}
{Yuan} F.,  {Narayan} R.,  2014, \mn@doi [\araa]
  {10.1146/annurev-astro-082812-141003}, \href
  {https://ui.adsabs.harvard.edu/abs/2014ARA&A..52..529Y} {52, 529}

\bibitem[\protect\citeauthoryear{Zaino et~al.,}{Zaino et~al.}{2020}]{Zaino2020}
Zaino A.,  et~al., 2020, \mn@doi [\mnras] {10.1093/mnras/staa107}, 492, 3872

\bibitem[\protect\citeauthoryear{Zoghbi, Miller  \& Cackett}{Zoghbi
  et~al.}{2019}]{Zoghbi2019}
Zoghbi A.,  Miller J.~M.,   Cackett E.,  2019, \mn@doi [\apj]
  {10.3847/1538-4357/AB3E31}, 884, 26

\bibitem[\protect\citeauthoryear{{de la Calle P{\'{e}}rez} et~al.,}{{de la
  Calle P{\'{e}}rez} et~al.}{2010}]{DelaCallePerez2010a}
{de la Calle P{\'{e}}rez} I.,  et~al., 2010, \mn@doi [\aap]
  {10.1051/0004-6361/200913798}, 524, A50

\bibitem[\protect\citeauthoryear{{in 't Zand} et~al.,}{{in 't Zand}
  et~al.}{2018}]{IntZand2018}
{in 't Zand} J. J.~M.,  et~al., 2018, \mn@doi [Sci. China Physics, Mech.
  Astron. 2018 622] {10.1007/S11433-017-9186-1}, 62, 1

\makeatother
\end{thebibliography}

\end{document}